\documentclass[twocolumn, aps, prb, 10pt, showpacs, english, preprintnumbers, amsmath, amssymb, superscriptaddress, longbibliography, nofootinbib]{revtex4-2}

\usepackage{comment}
\usepackage{graphicx}
\usepackage[dvipsnames]{xcolor}
\usepackage[normalem]{ulem}
\usepackage{appendix}
\usepackage{silence}
\usepackage[T1]{fontenc}
\usepackage{currvita}

\definecolor{orange}{rgb}{1,0.5,0}
\definecolor{goodgreen}{rgb}{0.1,0.5,0}
\definecolor{goodred}{rgb}{0.7,0,0}
\usepackage{lineno}

\usepackage[colorlinks,urlcolor=goodgreen,citecolor=blue,linkcolor=goodred]{hyperref}

\usepackage{float}
\graphicspath{ {figures/} }

\usepackage{babel}

\begin{document}

\title{Interplay between Superconductivity and Altermagnetism in Disordered Materials and Heterostructures}
\newcommand{\orcid}[1]{\href{https://orcid.org/#1}{\includegraphics[width=8pt]{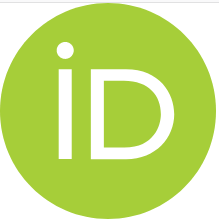}}}

\author{Rodrigo de las Heras\orcid{0009-0003-3640-1498}}
\email{rodrigo.delasheras@ehu.eus}
\affiliation{Centro de Física de Materiales (CFM-MPC) Centro Mixto CSIC-UPV/EHU, E-20018 Donostia-San Sebastián,  Spain}
\affiliation{Universidad del País Vasco (UPV/EHU)}

\author{Tim Kokkeler\orcid{0000-0001-8681-3376}}
\email{tim.h.kokkeler@jyu.fi}
\affiliation{Department of Physics and Nanoscience Center, University of Jyväskylä, P.O. Box 35 (YFL), FI-40014 University of Jyväskylä, Finland}

\author{Stefan Ili\'c\orcid{0000-0003-1406-9407}}
\affiliation{Department of Physics and Nanoscience Center, University of Jyväskylä, P.O. Box 35 (YFL), FI-40014 University of Jyväskylä, Finland}

\author{Ilya V. Tokatly\orcid{0000-0001-6288-0689}}
\affiliation{Donostia International Physics Center (DIPC), 20018 Donostia-San
Sebastián, Spain}
\affiliation{Departamento de Polimeros y Materiales
Avanzados, Universidad del Pais Vasco UPV/EHU, 20018 Donostia-San Sebastian,
Spain}
\affiliation{IKERBASQUE, Basque Foundation for Science, 48009 Bilbao, Spain}

\author{F. Sebastian Bergeret\orcid{0000-0001-6007-4878}}
\affiliation{Centro de Física de Materiales (CFM-MPC) Centro Mixto CSIC-UPV/EHU, E-20018 Donostia-San Sebastián,  Spain}
\affiliation{Donostia International Physics Center (DIPC), 20018 Donostia-San Sebastián, Spain}

\begin{abstract}
We study the interplay between superconductivity and altermagnetism in disordered systems using recently derived quantum kinetic transport equations.
Starting from this framework, we derive the Ginzburg–Landau free energy and identify, in addition to the conventional pair-breaking term, a coupling between the spin and the spatial variation of the superconducting order parameter. Two distinct effects emerge from this coupling. The first is a nonlinear magnetoelectric effect, in which a supercurrent ({\it i.e.}, a phase gradient) induces a spin texture; this contribution is quadratic in the phase gradient. The second effect arises when the magnitude, rather than the phase, of the superconducting order parameter varies in space, likewise leading to a finite magnetization.
We show that these two contributions compete in the case of an Abrikosov vortex, where both the amplitude and phase of the order parameter vary spatially. The effect associated with amplitude variations also gives rise to a proximity-induced magnetization (PIM) in hybrid structures composed of a superconductor (S) and an altermagnet (AM). Using quasiclassical theory, we analyze the PIM in diffusive S/AM bilayers and S/AM/S Josephson junctions, and determine the induced magnetization profiles. In Josephson junctions, where both the PIM and the nonlinear magnetoelectric effect coexist, we further predict the occurrence of $0$–$\pi$ transitions.
\end{abstract}
\maketitle
\section{Introduction}
A recent development in condensed matter physics that attracts considerable attention is the prediction of altermagnets \cite{vsmejkal2022emerging,vsmejkal2022beyond}, a class of materials combining antiferromagnetic type of spin distribution (zero net magnetization per unit cell) with spin–split electronic bands. This unique combination makes them promising candidates for spintronics applications, where they can exhibit intriguing transport phenomena \cite{zarzuela2025transport, vsmejkal2022giant,reichlova2024observation,dou2025anisotropic,sun2025tunneling,cjzw-j4v7}.

An equally interesting research direction is the combination of altermagnetism and superconductivity~\cite{lu2024varphi,hu2025nonlinear,alam2026proximityinducedsuperconductivityemergingtopological,zyuzin2024magnetoelectric,chourasia2025thermodynamic,sharma2025tunable,fukaya2025josephson,fukaya2025superconducting,banerjee2024altermagnetic,papaj2023andreev,chakraborty2024zero,sun2023andreev,ouassou2023dc,bobkov2025inverse,PhysRevB.111.L121401,prnx-47mk,li2025spinpolarizedjosephsonsupercurrentnodeless,8cwk-nynj}, both in coexisting situations and in hybrid structures. Most of these works focus on purely ballistic systems. However, a recent work~\cite{TimNLSM} derives the kinetic equations describing transport in normal and superconducting diffusive systems with altermagnets. These equations were obtained from the nonlinear sigma model (NLSM) developed in Ref.~\cite{TimNLSM} using a symmetry-based approach \cite{kokkeler2025universal}.

In this work, we use the framework of Ref.~\cite{TimNLSM} to explore the interplay between superconductivity and altermagnetism in diffusive systems. First, we derive from the NLSM the Ginzburg–Landau (GL) free-energy functional for a disordered superconducting altermagnet. The simple structure of this functional reveals that, besides the exchange-field–induced pair-breaking mechanisms that renormalize the usual GL coefficients, a new term emerges, associated with the altermagnetic order. This term is of the second order in the gradients of the order parameter and is closely related to the phenomenon of piezomagnetism \cite{dzialoshinskii1958problem}.

It is known that $d$--wave altermagnets allow for a linear piezomagnetic effect \cite{schiff2025collinear,khodas2025tuning}, that is, a linear relation $M_i=\alpha_{ijk}u_{jk}$ between the magnetization $M_i$ and the strain tensor $u_{jk}$. In other words, the symmetry of altermagnets implies the existence of a third rank tensor $\alpha_{ijk}$ which is odd under time reversal and symmetric in its last two indices. An immediate consequence is that, in the superconducting state, such materials may exhibit a magnetization induced  by inhomogeneities of the order parameter, $M_i\sim \alpha_{ijk}(\partial_j\Delta)(\partial_k\Delta^*)$. As we show below, the GL functional derived from the NLSM indeed contains precisely such a term, arising naturally from the symmetry-based approach of Ref.~\cite{TimNLSM}.

This gradient term contains two distinct physical mechanisms. First, the magnetoelectric effect predicted in \cite{hu2025nonlinear,zyuzin2024magnetoelectric}, in which a phase gradient, {\it i.e.}, a supercurrent, induces a magnetic moment. Symmetry dictates that this effect appears only at second order in the phase gradient. Second, our analysis shows that the same term also describes a distinct effect in which a finite magnetization arises when the \emph{magnitude}, rather than the \emph{phase}, of the order parameter varies in space. Such an effect should manifest in materials where superconductivity and altermagnetism coexist and the order parameter is real and spatially inhomogeneous, as in a Larkin--Ovchinnikov--type phase.

Moreover, the latter effect is also expected in a hybrid S/AM bilayers, where superconducting correlations penetrate into the altermagnet via the proximity effect, decaying away from the S/AM interface and thereby generating a finite magnetic moment even in the absence of an injected supercurrent. In this geometry, the presence of a normal vector to the S/AM interface allows this effect to appear already to first order in the magnitude gradient. We refer to this mechanism as a proximity-induced magnetization (PIM). The generated magnetization is an emergent property of the hybrid junction, since both the superconductor and the altermagnet intrinsically lack a net magnetic moment.

With the help of the Usadel equation for altermagnets, we present an exhaustive analysis of the proximity-induced magnetization (PIM), first in an S/AM bilayer. We derive the characteristic length of the superconducting proximity effect in altermagnets, which depends on both the altermagnetic parameter $K$ and the pair-breaking rate $\Gamma$. As in ferromagnets~\cite{buzdin1991josephson}, the superconducting condensate not only decays inside the AM but also oscillates in space. 
Within the same framework, we also analyze the PIM in an S/AM/S Josephson junction and determine the critical current as a function of the material parameters. When a supercurrent is induced in the junction, the PIM coexists with the magnetoelectric effect, which is quadratic in the phase difference between the superconducting leads. We also find that $0$–$\pi$ transitions are possible in such systems, for example, upon varying the temperature.

Our analysis is based on general equations valid for any microscopic realization of an altermagnet. As an example we also include a specific effective low-energy Hamiltonian describing a disordered system with coexisting superconducting and altermagnetic order. From this Hamiltonian, we derive in Appendix \ref{microscopic} the quasiclassical equations in the diffusive limit and establish the connection between the effective parameters of the nonlinear sigma model.

The paper is organized as follows. In Section~\ref{sec2}, we present the GL free energy for a superconducting altermagnet and highlight the main effects arising from the interplay between superconducting and altermagnetic orders. In Section~\ref{S/AMhetero}, we analyze the proximity-induced magnetization (PIM) in S/AM and S/AM/S structures. For S/AM/S Josephson junctions, we show that the magnetization arises both from the supercurrent and from spatial variations of the modulus of the superconducting condensate, and we discuss the possibility of $0$–$\pi$ transitions in the Josephson current. In Section~\ref{Microscopicderiv}, we express the parameters of the altermagnet in terms of an effective low-energy model describing altermagnetic order. Finally, we summarize our conclusions in Section~\ref{sec:Conclusions}.

\section{Ginzburg–Landau Free Energy}\label{sec2}
In this section we derive the 
GL free energy for a dirty superconducting altermagnet. Our starting point is the expression for the free energy functional \cite{virtanen2020quasiclassical}:
\begin{equation}
    F = \frac{\nu_0}{\lambda}|\Delta|^2 - T \sum_{n=-\infty}^{\infty} i S[\check{g}, \omega_n]\;, 
    \label{eq:free_energy}
\end{equation}
where here $\lambda$ is the phonon-mediated pairing interaction constant, $\nu_0$  the density of states per spin at the Fermi level, $T$ the temperature and $\Delta$ the superconducting order parameter. 
The second term in Eq.~\eqref{eq:free_energy}, including terms due to superconductivity and altermagnetism, reads \cite{TimNLSM}:
\begin{align}
    i S[\check{g}, \omega_n] &= \frac{\pi \nu_0}{2} \operatorname{Tr} \bigg( -\frac{D}{4} (\check\nabla \check{g})^2 + \omega_n \check{\tau}_3 \check{g} + \check{\Delta} \check{g} \notag\\
    &\quad + \frac{1}{4} \Gamma_{ab} \check{\tau}_3 \hat{\sigma}_a \check{g} \check{\tau}_3 \hat{\sigma}_b \check{g} - \frac{D}{4} T_{ajk} \hat{\sigma}_a \check{\tau}_3 \check\nabla_j \check{g} \check\nabla_k \check{g} \notag\\
    &\quad - \frac{iD}{4} K_{ajk} \hat{\sigma}_a \check{\tau}_3 \check{g} \check\nabla_j \check{g} \check\nabla_k \check{g} +ih_a\hat\sigma_a\check\tau_3\check g\bigg)\;.
    \label{ActionAltermagnet}
\end{align}
The above expression corresponds to the Luttinger–Ward functional in terms of the quasiclassical Green's function $\check{g}$ \cite{virtanen2020quasiclassical}. In fact, it is the saddle point value of the effective action of the NLSM for an equilibrium superconductor\footnote{Technically, the connection to the free energy is obtained assuming that the time integration in the NLSM goes over the Matsubara imaginary track.}. It has a $4\times 4$ matrix structure in Nambu–spin space \cite{bergeret2005odd}, and satisfies the normalization condition $\check g^2=1$. The Pauli matrices spanning the Nambu and the spin subspaces are denoted by $\check{\tau}_i$ and $\hat{\sigma}_i$, respectively.

The specific structure of the tensors in Eq.~\eqref{ActionAltermagnet} is determined purely by symmetry, in the same spirit as the construction of the nonlinear sigma model (NLSM) recently proposed in Refs.~\cite{TimNLSM,kokkeler2025universal}.  

In Eq.~\eqref{ActionAltermagnet}, $D$ is the diffusion coefficient, $\omega_n=\pi T(1+2n)$  the Matsubara frequencies, $\check\Delta=|\Delta|e^{i\check\tau_3\varphi}\check \tau_1$ the matrix pair potential and $\varphi$ the phase of the superconductor. The covariant derivative is defined as $\check\nabla_k\check Q=\partial_k\check Q-iA_k[\check \tau_3,\check Q]$, where $A_k$ is the vector potential. In the last term, $h_a = g\mu_B B_a + h_a^{ex}$ stands for the total spin-splitting field that accounts for the Zeeman splitting due to the magnetic field $B_i=\varepsilon_{ijk}\partial_jA_k$ and, possibly, an extrinsic exchange field $h_a^{ex}$. Here, $\mu_B$ is the Bohr magneton and $g$ is the Landé $g$-factor. Formally, $h_a^{ex}$ appears as a field conjugated to the spin density $S_a$. The spin can thus be computed by differentiating the free energy with respect to  the spin-splitting  field,
$S_a = -\frac{\partial F}{\partial h_a^{ex}} = -\frac{\partial F}{\partial h_a}$.

The tensors $\Gamma_{ab}$, $K_{ajk}$, and $T_{ajk}$ describe different aspects of exchange interactions in magnetic materials. The tensor $\Gamma_{ab}$ represents a spin-relaxation term associated with magnetic interactions, and is present in ferromagnets, antiferromagnets and altermagnets. The tensors $K_{ajk}$ and $T_{ajk}$, are odd under time-reversal symmetry and symmetric in the last two indices. They describe the physics of spin-dependent transport phenomena \cite{TimNLSM}. Specifically, $T_{ajk}$ is related  to the difference in diffusion constant for electrons with opposite spin along the collinear axis, while $K_{ajk}$ to the Larmor precession in the presence of gradients \cite{TimNLSM}. Consequently, $T_{ajk}$ and $K_{ajk}$ only appear in systems with spin-split bands, such as ferromagnets and altermagnets, but vanish in antiferromagnets, where the absence of spin splitting forbids them by symmetry.

A key distinction between ferromagnets and altermagnets is that in ferromagnets there is an intrinsic exchange field $h^{int}_a$ responsible for the ferromagnetic spin splitting, while in altermagnets the intrinsic exchange field is zero. It can appear only when applied externally, or as formal source field used to define the spin density from the derivative of the free energy. Hence, spin phenomena in ferromagnets are primarily governed by $h_a^{int}$, while in altermagnets the dominant spin-related effects originate from the tensors $K_{ajk}$ and $T_{ajk}$.

Minimization of the energy Eq.~\eqref{eq:free_energy} with respect to $\check g$ gives the Usadel equation:
\begin{align}\label{UsadelAltermagnet}
    \check\nabla_k\check{\mathcal J}_k=&[\check g,(\omega_n+ih_a\hat\sigma_a)\check\tau_3+\check\Delta]
    \notag\\
    &+\frac{1}{2}\Gamma_{ab}[\check g,\check{\tau}_3\hat\sigma_a\check g\check{\tau}_3\hat\sigma_b]+\check{\mathcal{T}}\;,
\end{align}
where summation over repeated indices is implied. The matrix current $\mathcal{\check{J}}_k$ and torque $\mathcal{\check{T}}$ are position and Matsubara frequency dependent matrices in Nambu-spin space, given by:
\begin{align}\label{Eq:matrixJ}
    \check{\mathcal{J}}_k=&-D\bigg(\check{g}\check\nabla_k\check{g}+\frac{1}{4}T_{ajk}\{\check{\tau}_3\hat \sigma_a+\check{g}\check{\tau}_3\hat \sigma_a\check{g},\check{g}\check\nabla_j\check{g}\}\notag\\
    &+\frac{i}{4}K_{ajk}[\check{\tau}_3\hat \sigma_a+\check{g}\check{\tau}_3\hat \sigma_a\check{g},\check\nabla_j \check g] \bigg)
\\ \label{Eq:matrixT}
    \check{\mathcal T}=& D\bigg(\frac{1}{4}T_{ajk}[\check{\tau}_3\hat \sigma_a, \check\nabla_j\check{g} \check\nabla_k \check{g}]+\frac{i}{4}K_{ajk}[\check{\tau}_3\hat \sigma_a,\check{g}\check\nabla_j\check{g} \check\nabla_k \check{g}]\bigg).
\end{align}
It is important to emphasize that the Usadel equation, Eq.~\eqref{UsadelAltermagnet}, is formulated solely on the basis of symmetry considerations. As a result, all symmetry-allowed terms appear as independent phenomenological coefficients.  In Section~\ref{Microscopicderiv}, we derive these coefficients from a specific  low-energy Hamiltonian, thereby confirming its general structure and providing expressions for the coefficients in terms of the parameters of  that particular realization.

To derive the GL free energy we consider a bulk superconductor with altermagnetic order near its critical temperature $T_c$. Close to $T_c$ the Usadel equation Eq.~\eqref{UsadelAltermagnet} can be linearized \cite{bergeret2005odd}, and the Green’s function parametrized as:
\begin{equation}\label{parametrization}
     \check g=\begin{pmatrix}
\operatorname{sgn}(\omega_n)(1-\frac{1}{2}\hat f\hat{\tilde f}) & \hat f \\
\hat{\tilde{f}} & -\operatorname{sgn}(\omega_n)(1-\frac{1}{2}\hat{\tilde f}\hat f) \\
\end{pmatrix}\;,
\end{equation}
where $\hat f$ and $\hat{\tilde f}$ are the anomalous Green’s functions in spin space that describe the condensate.   They decompose as:
\begin{equation}
\hat f=f_s+f_t^a\hat \sigma_a\;,\quad
\hat{\tilde{f}}=\tilde f_s+\tilde f_t^a\hat\sigma_a\;,
\end{equation}
where $f_s$ is the singlet component of the condensate and $f_t^a$ the triplet components.

Substituting this parametrization into the Usadel equation, assuming that only $\Gamma_{zz}\equiv\Gamma\neq0$ (which is motivated in section~\ref{Microscopicderiv}), neglecting spatial derivatives (bulk limit) and to linear order in $\Delta$ and $h_a$, one finds:
\begin{align}
    &f_s=\frac{\Delta}{|\omega_n|+\Gamma}\;,\quad \quad\qquad\ \tilde{f_s}=\frac{\Delta^*}{|\omega_n|+\Gamma}\;,
\\
    &f_t^a=-\frac{i\operatorname{sgn}(\omega_n)\Delta}{(|\omega_n|+\Gamma)^2}h_a\;,\quad \tilde f_t^a=-\frac{i\operatorname{sgn}(\omega_n)\Delta^*}{(|\omega_n|+\Gamma)^2}h_a\;.
\end{align}
Inserting these solutions into Eq.~\eqref{eq:free_energy}, and keeping terms up to  second  order in $|\Delta|$ and linear order in $h_a$, gives the GL free energy functional:
\begin{align}\label{GLFreeEnergy}
F=&F_n+a|\Delta|^2+c|\Pi\Delta|^2+c'h_aK_{ajk}\Pi_j\Delta\Pi_k^*\Delta^*\;,
\end{align}
where $\Pi_j=-i\partial_j-2A_j$, $F_n$ is the free energy of the normal state and:
\begin{equation}\label{eq:a}
    a=\frac{\nu_0}{\lambda}-2\pi\nu_0 T\sum_{n=0}^{n_{max}}\frac{1}{\omega_n+\Gamma}\;,
\end{equation}
\begin{equation}
    c=\pi\nu_0 T\sum_{n=0}^{\infty}\frac{D}{(\omega_n+\Gamma)^2}\;,
\end{equation}
\begin{equation}\label{eq:cjk}
    c'=\pi \nu_{0} T\sum_{n=0}^{\infty}\frac{2D}{(\omega_n+\Gamma)^3}\;,
\end{equation}
where in Eq.~\eqref{eq:a}, we introduce the Debye frequency as a cutoff $n_{max}=\omega_D/(2\pi T)$.

The resulting free energy coincides in structure with the GL functional presented in Ref.~\cite{zyuzin2024magnetoelectric}. However, our derivation establishes a connection between the corresponding GL coefficients and the parameters of the higher level Usadel theory that is valid for any temperature.

Note that the tensor $T_{ajk}$ in Eqs.~(\ref{UsadelAltermagnet}-\ref{Eq:matrixT}) does not enter the free energy, {\it i.e.}, it does not 
contribute to equilibrium properties in single domain superconducting collinear altermagnets, in agreement with Ref.~\cite{TimNLSM}.

At $T = T_c$, the coefficient $a$ must vanish, and therefore:
\begin{equation}\label{lambdabegin}
\frac{1}{\lambda} = 2\pi T_c \sum_{n=0}^{n_{max}} \frac{1}{\omega_n + \Gamma}\ .
\end{equation}
By subtracting the equivalent equation for zero $\Gamma$, one obtains the expression determining the critical temperature:
\begin{align}\label{Selfconsistency}
    \ln\!\left(\frac{T_c}{T_{c0}}\right)
    &= 2\pi T_c \sum_{n=0}^\infty \left( \frac{1}{\omega_n + \Gamma} - \frac{1}{\omega_n} \right) \notag\\
    &= \psi^{(0)}\!\left(\frac{1}{2}\right)
     - \psi^{(0)}\!\left(\frac{1}{2} + \frac{\Gamma}{2\pi T_c}\right),
\end{align}
where $\psi^{(0)}(z)$ is the digamma function and $T_{c0}$ is the critical temperature in the absence of  $\Gamma$.
Equation~\eqref{Selfconsistency} is formally identical to the Abrikosov–Gor'kov theory of magnetic impurities in superconductors~\cite{abrikosov2017fundamentals}, with $\Gamma$ being the pair-breaking parameter.

More interesting is the last term in the first line of Eq.~(\ref{GLFreeEnergy}), proportional to the tensor $K_{ajk}$. This term couples gradients of the order parameter to the spin-splitting field $h_a$ and therefore to the spin polarization. As a result, it gives rise to a finite magnetization, which can be obtained by differentiating the free energy with respect to $h_a$ \cite{white2007quantum}:
\begin{align}\label{MagnetizationGL}
    M_a&=-g\mu_B\frac{\partial F}{\partial h_a}=-\sum_{n=0}^{\infty}\frac{2g\mu_B\nu_0\pi TD}{(\omega_n+\Gamma)^3}K_{ajk}\Pi_j\Delta\Pi_k^*\Delta^*\notag\\
    &=\frac{g\mu_B\nu_0 D}{8\pi^2T^2}\psi^{(2)}\left(\frac{1}{2}+\frac{\Gamma}{2\pi T}\right)K_{ajk}\Pi_j\Delta\Pi_k^*\Delta^*\;,
\end{align}
where $\psi^{(2)}(z)$ is the tetragamma function \cite{abramowitz2006handbook}. 
 The induced magnetization is proportional to the tensor $K_{ajk}$ and decreases with increasing $\Gamma$, since $\psi^{(2)}(z)$ is a monotonically decreasing function for $z>0$. It is worth noting that the existence of the term proportional to the $K$-tensor  in the GL functional, Eq.~(\ref{GLFreeEnergy}), for a superconductor with altermagnetic order can also be justified solely on the basis of symmetry arguments.

Eq.~\eqref{MagnetizationGL} predicts the appearance of a finite magnetization arising from the spatial variation of the order parameter $\Delta(\mathbf{r})$. One can distinguish two situations. The first arises when $|\Delta|$ is constant in space but we have a finite phase gradient $\partial_k\varphi$. This describes the magnetoelectric effect, previously discussed in Refs.~\cite{zyuzin2024magnetoelectric,hu2025nonlinear}, corresponding to a magnetic moment induced by a supercurrent. Importantly, unlike the conventional Edelstein effect in superconductors with spin–orbit coupling~\cite{edelstein1995magnetoelectric}, this response is nonlinear in the phase gradient;  the induced magnetization scales quadratically with the supercurrent. For this reason, it is referred to as the nonlinear superconducting magnetoelectric effect
 \cite{hu2025nonlinear,
zyuzinarXIVE2024magnetoelectric} \footnote{Originally, in the published Ref.~\cite{zyuzin2024magnetoelectric} a magnetoelectric effect linear in the phase gradient was suggested. The corresponding equations have later been corrected, see the latest arXiv version \cite{zyuzinarXIVE2024magnetoelectric}. A similar linear magnetoelectric effect was reported in \cite{giil2024quasiclassical}, which however appeared due to the use of an improper boundary condition. From simple symmetry arguments, see Eq.~(\ref{GLFreeEnergy}), a linear coupling between the phase gradient and the magnetization is not possible in superconducting altermagnets under equilibrium conditions.
 }.

The second situation, described by Eq.~\eqref{MagnetizationGL}, corresponds to the appearance of a magnetization in the presence of spatial variations of the magnitude $\partial_k|\Delta|$, but not of its phase. In other words, even in the absence of any supercurrent, a magnetization may be induced. Such variations occur, for example, in inhomogeneous superconductors with a real order parameter, as in Larkin–Ovchinnikov phase \cite{larkin1965nonuniform}.

\begin{figure}[tb]
    \centering
    \includegraphics[width=\linewidth]{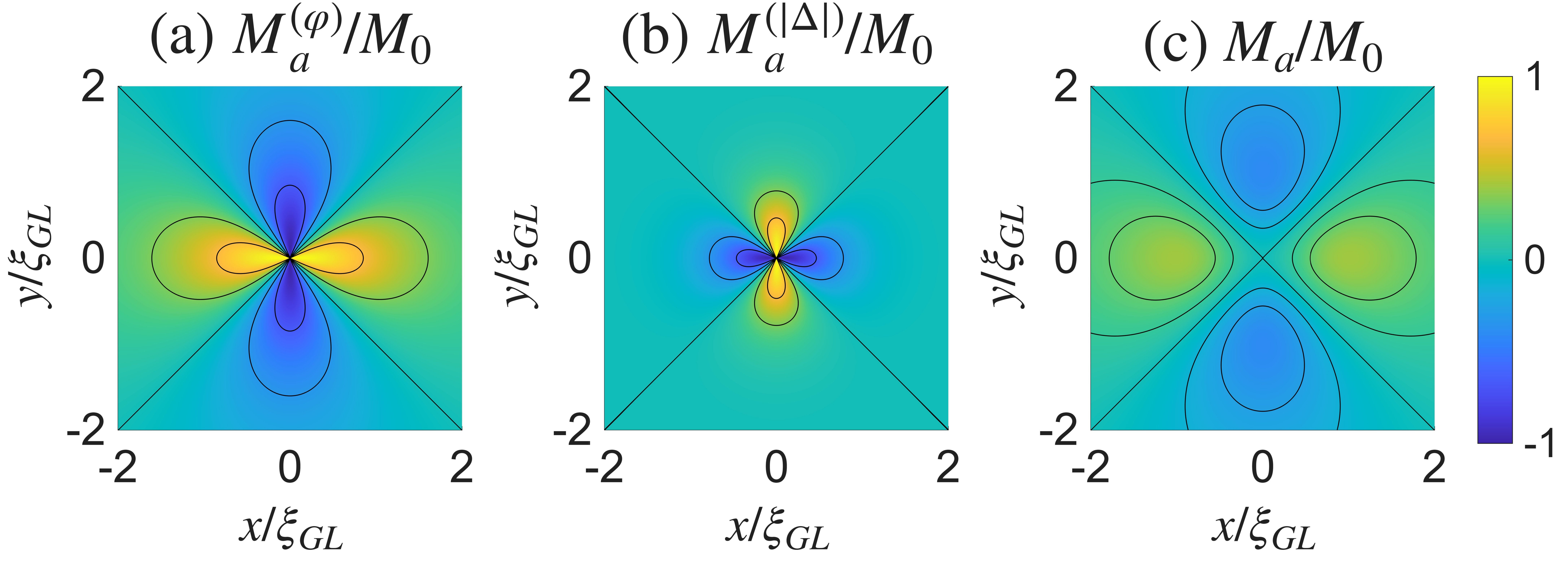}
    \caption{Magnetization induced in the presence of an Abrikosov vortex in a superconducting altermagnet with vorticity $m=1$. Contributions due to (a)   the nonlinear magnetoelectric effect, and (b) due to the variation of $|\Delta|$.  (c)  Total magnetization $M_a = M_a^{(\varphi)} + M_a^{(|\Delta|)}$. The underlying $d$-wave symmetry of the altermagnet is manifested as four lobes with alternating sign. } 
    \label{vortex}
\end{figure}

An interesting example that illustrates both types of effects, {\it i.e.}, the induced magnetization due to the nonlinear magnetoelectric effect and that arising from an inhomogeneous order parameter, is the case of an Abrikosov vortex. 
For the single-quantum vortex, with vorticity $m=1$, the order parameter can be approximated by
$\Delta(\mathbf{r}) \approx |\Delta_{\infty}|\tanh\left({r}/{\xi_{GL}}\right)e^{i \varphi}$ \cite{tinkham2004introduction},
where $r = \sqrt{x^2 + y^2}$, $\varphi = \arctan(y/x)$, $\xi_{\text{GL}}$ is the coherence length and $\Delta_\infty$ is the order parameter in the bulk. 
We choose the crystallographic axes of the altermagnet such that $K_{axx} = -K_{ayy} = K$ \cite{TimNLSM}, and all other components of the tensor $K$ are zero. 
The nonlinear magnetoelectric contribution stems from the circulating supercurrents around the vortex, and is given by:
\begin{align}
\frac{M_a^{(\varphi)}}{M_0} &= -\xi_{\text{GL}}^2 \big[(\partial_x \varphi)^2 - (\partial_y \varphi)^2\big] \tanh^{2}\!\left(\frac{r}{\xi_{\text{GL}}}\right)
\notag\\
&= \frac{\xi_{\text{GL}}^2}{r^2}\cos(2\varphi)\tanh^{2}\!\left(\frac{r}{\xi_{\text{GL}}}\right)\;,
\end{align}
where $M_0=\frac{g\mu_B\nu_0D|\Delta_\infty|^2}{8\pi^2T^2\xi_{\text{GL}}^2}K|\psi^{(2)}\left(\frac{1}{2}+\frac{\Gamma}{2\pi T}\right)|$. The second contribution from the variation of $|\Delta|$ is:
\begin{align}
\frac{M_a^{(|\Delta|)}}{M_0} &=-\frac{\xi_{\text{GL}}^2}{|\Delta_\infty|^2} \left[ (\partial_x |\Delta|)^2 - (\partial_y |\Delta|)^2 \right] \notag\\
  &=  -\frac{\cos(2\varphi)}{\cosh^4\left(\frac{r}{\xi_{\text{GL}}}\right)}\;.
\end{align}
Interestingly, both contributions vanish along the nodal line of the altermagnet $|x| = |y|$, reflecting its $d$-wave symmetry. Moreover, the two contributions have opposite signs and, in amplitude, decay monotonically away from the vortex core, see Fig.~\ref{vortex}\textcolor{goodred}{(a)--(b)}. The total magnetization $M = M_a^{(\varphi)} + M_a^{(|\Delta|)}$, exhibits a maximum at distances of the order of $\xi_{\text{GL}}$ from the vortex core, see Fig.~\ref{vortex}\textcolor{goodred}{(c)}. 

We can also evaluate the supercurrent to identify the contribution arising from the $K_{ajk}$ term:
\begin{align}
  j_k=-\frac{\partial F}{\partial A_k}=&2\Big[c(\Delta\Pi_k^*\Delta^*+\Delta^*\Pi_k\Delta)
  \notag\\
  &+c'h_a K_{ajk}(\Delta\Pi_j^*\Delta^*+\Delta^*\Pi_j\Delta)\Big]\;.
\end{align}

This expression shows that, depending on the crystal orientation of the altermagnet, the term proportional to $h_a K_{ajk}$ can generate either a parallel or a perpendicular component of the supercurrent.
For a phase gradient along the $x$ direction and the orientation $K_{axx} = -K_{ayy} = K$, the induced contribution is parallel to the current. Notably, since this term is odd in $h_a$, the resulting current becomes asymmetric such that $j_x(h_a) \neq j_x(-h_a)$. In contrast, for $K_{axy} = K_{ayx} = K$, a transverse component $j_y$ arises, as discussed in Ref.~\cite{zyuzin2024magnetoelectric}.

Next to this, in the presence of an inhomogeneity of $\Delta$, there is a free energy difference for opposite $K_{ajk}$:
\begin{align}
    F(K_{ajk})-F(-K_{ajk}) = 2c'h_a K_{ajk}\Pi_j\Delta \Pi_k^{*}\Delta^{*}\;.
\end{align}
For example, in the presence of a phase gradient $q$, this expression reads $F(K_{ajk})-F(-K_{ajk}) = 2c'h_a K_{ajk}q^2|\Delta|^2$.
The free energy difference is linear in $K_{ajk}$, {\it i.e.}, first order in the N\'eel vector, in a similar manner in which strain can introduce a term first order in the N\'eel vector \cite{dzialoshinskii1958problem,aoyama2024piezomagnetic,khodas2025tuning,schiff2025collinear}. This shows that supercurrents, as well as magnitude gradients, can be used to aid the switching of the N\'eel vector of altermagnets using a magnetic field $B_a$ that will induce the spin-splitting field $h_a$.

All these effects, so far considered for bulk superconducting altermagnets, can also appear in hybrid S/AM structures. 
In this case, inversion symmetry is broken, and there exists a polar vector normal to the interface, $n_k$. We can then construct a term, similar to the third term on the first line of Eq.~(\ref{GLFreeEnergy}), but now one of the spatial derivatives of the order parameter will be replaced by $n_k$, which may lead to a spatially dependent magnetic moment appearing in the AM. 
This is the focus of the next sections, where we analyze the proximity induced magnetization for S/AM and S/AM/S heterostructures beyond the GL limit, by solving the Usadel equation. 

\section{Superconductor/Altermagnet Heterostructures}\label{S/AMhetero}

In this section we consider hybrid  S/AM  and S/AM/S structures. 
We focus on a two-dimensional, single domain, $d$-wave collinear altermagnet. We denote the collinear axis as the $z$-direction in spin space ($a = 3$). In this case, the most general form of the $K_{ajk}$ tensor is \cite{TimNLSM}:
\begin{equation}
  K_{3jk}=\begin{pmatrix}
K\cos(2\alpha) & K\sin(2\alpha)\\
K\sin(2\alpha) & -K\cos(2\alpha)
\end{pmatrix}\label{eq:Kparameterization}\;,
\end{equation}
where $\alpha$ denotes the angle between the $x$ axis and the crystallographic axis. Thus, we simplify the notation by introducing $K_{jk}\equiv K_{3jk}$. We consider that only one component of the spin-relaxation tensor $\Gamma\equiv\Gamma_{33}$ does not vanish, as this is the case in the low-energy model of Sec.~\ref{Microscopicderiv}.

To investigate the proximity effect we need to solve Eq.~\eqref{UsadelAltermagnet}. At the interface between the altermagnet and the superconductor, we impose the extended Kupriyanov–Lukichev boundary conditions \cite{kuprianov1988influence,belzig1999quasiclassical,TimNLSM}:
\begin{equation}
\label{eq:KL-BC}
n_k\check{\mathcal{J}}_k=\frac{D}{2\xi_0\gamma_B}[\check g_S,\check g]\;,
\end{equation}
where $\check g_S$ is the Green's function of the superconductor, $\check{\mathcal J}$ is defined in Eq.~\eqref{Eq:matrixJ}, $n_k$ is the unit normal vector pointing outward from the altermagnet, $\Delta_0=|\Delta(T=0)|$ is the zero temperature pair potential and $\xi_0=\sqrt{D/\Delta_0}$ is the coherence length in the superconductor. For simplicity, we assume that the diffusive constant $D$ is the same in both the superconductor and the altermagnet throughout this work. The constant $\gamma_B$ characterizes the interface transparency:
\begin{equation}
    \gamma_B=\frac{R_b\sigma_n}{\xi_0}\;,
\end{equation}
with $R_b$ the interface resistance per unit area and $\sigma_n=2e^2D\nu_0$ the Drude conductivity. At an interface with vacuum $\gamma_B\to\infty$.

We assume  that the superconductor is not affected by the inverse proximity effect, and hence is described by the bulk BCS Green’s function \cite{bergeret2005odd}:
\begin{equation}
    \check g_S\approx\check{g}_{BCS}=\frac{1}{\sqrt{\omega_n^2+|\Delta|^2}}(\omega_n\check \tau_3+\check\Delta)\;.
\end{equation}
Furthermore, we assume a weak proximity effect, such that  the Usadel equation in the AM can be linearized \cite{bergeret2005odd}, as in Eq.~\eqref{parametrization}. For a collinear axis along $z$, the pair amplitudes $\hat f$ and $\hat{\tilde f}$ are diagonal in spin space:
\begin{equation}\label{parametrizationpairs}
\hat f=\begin{pmatrix}
f_+ & 0 \\
0 & f_- \\
\end{pmatrix},\quad
\hat{\tilde{f}}=\begin{pmatrix}
\tilde f_+ & 0 \\
0 & \tilde f_- \\
\end{pmatrix}\;.
\end{equation}
Substituting this parametrization into the Usadel equation Eq.~\eqref{UsadelAltermagnet} with $\check \Delta=0$ (normal state altermagnet) and neglecting nonlinear terms in the pair amplitudes yields:
\begin{align}\label{LinearUsadel}
    (1\pm i\operatorname{sgn}(\omega_n)K\cos(2\alpha))\partial_x^2f_{\pm}\pm2i\operatorname{sgn}(\omega_n)K\sin(2\alpha)\partial_x\partial_yf_{\pm} \notag\\
    +(1\mp i\operatorname{sgn}(\omega_n)K\cos(2\alpha))\partial_y^2f_{\pm}=\frac{2(|\omega_n|+\Gamma)}{D}f_{\pm}\;,
\end{align}
and analogous equations are obtained for $\tilde f_{\pm}$ by replacing $f_{\pm}\leftrightarrow\tilde f_{\pm}$. Note that, as in the GL case, the tensor $T_{ajk}$ does not enter the equations because we consider an equilibrium, collinear system. In noncollinear systems, or under nonequilibrium conditions, the tensor $T_{ajk}$ can give rise to spin-filtering and spin-splitting effects \cite{sigales2026spinsplitterinverseeffects,TimNLSM}.

In the following two subsections we solve Eq.~\eqref{LinearUsadel}, together with boundary conditions obtained by linearization of Eq.~\eqref{eq:KL-BC}  for the S/AM bilayer and the S/AM/S Josephson junction (see Appendix \ref{Ap:BoundaryCons} for details). 

\subsection{S/AM bilayer}\label{sec:bilayer}
\begin{figure}[H]
    \centering
    \includegraphics[width=0.9\linewidth]{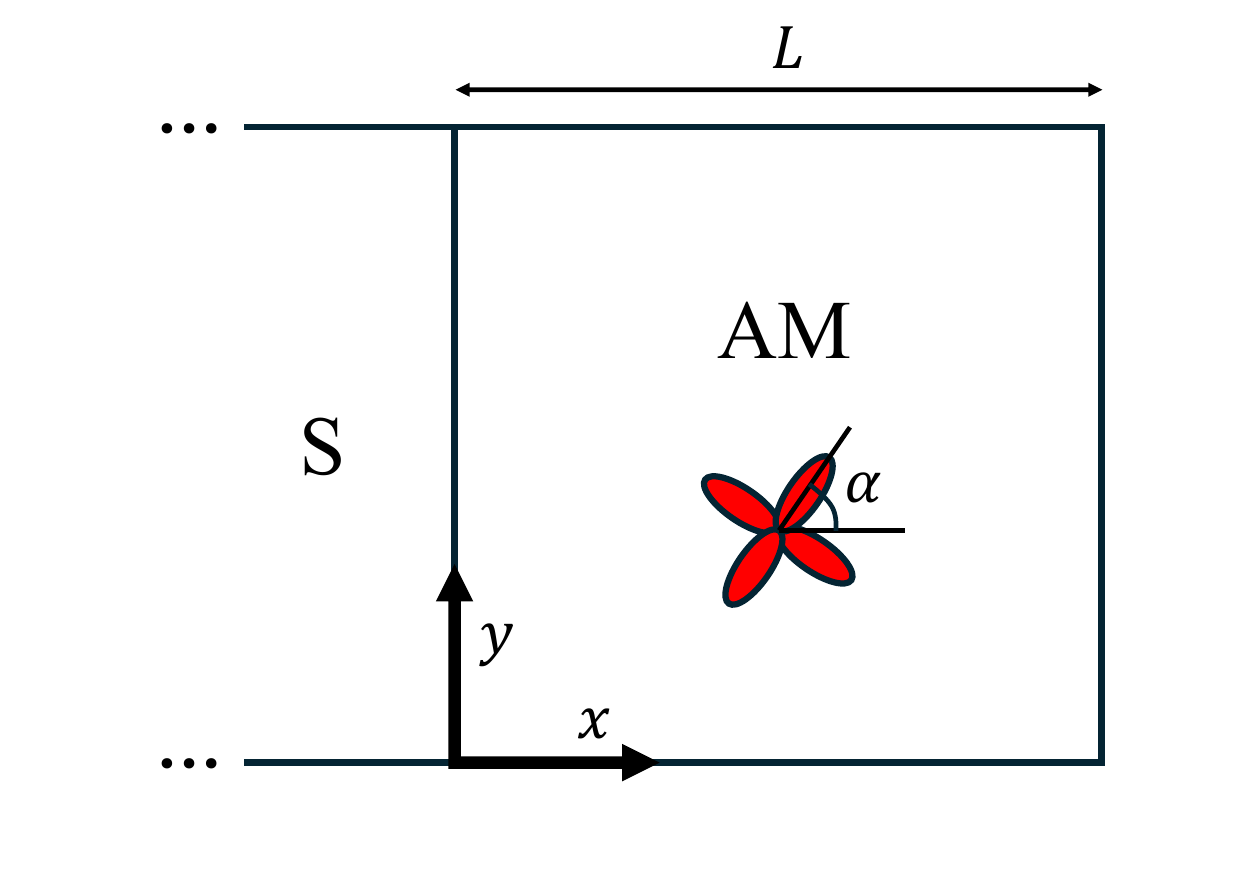}
    \caption{A square altermagnet of dimensions $L\times L$ next to an infinite (bulk) superconductor at $x=0$. The remaining edges of the altermagnet ($x=L$, $y=0$ and $y=L$) are bounded by vacuum. We draw a four-lobed red flower, reminiscent of $d$-wave symmetry, to represent the angle $\alpha$ between the $x$ axis and the crystallographic axis of the altermagnet.} 
    \label{SAM_sketch}
\end{figure}

We consider a square altermagnet of side $L$ in contact with a bulk superconductor, as illustrated in Fig.~\ref{SAM_sketch}.

For an arbitrary angle $\alpha$,  Eq.~\eqref{LinearUsadel} must be solved numerically. However, for $\alpha=0$ the problem becomes independent of the coordinate $y$, allowing for an analytic solution of the pair amplitudes:
\begin{equation}\label{pairs}
f_\pm=\frac{|\Delta|\cosh(\kappa_{\omega_n}^\pm(x-L))}{\gamma_B\xi_0\sqrt{\omega_n^2+|\Delta|^2}(1\pm i\operatorname{sgn}(\omega_n)K)\kappa_{\omega_n}^\pm \sinh(\kappa_{\omega_n}^\pm L)}
\end{equation}
\begin{equation}
\kappa_{\omega_n}^\pm=\sqrt{\frac{2(|\omega_n|+\Gamma)}{D(1\pm i\operatorname{sgn}(\omega_n)K)}}\;.
\end{equation}

\begin{figure}[H]
    \centering
    \includegraphics[width=0.9\linewidth]{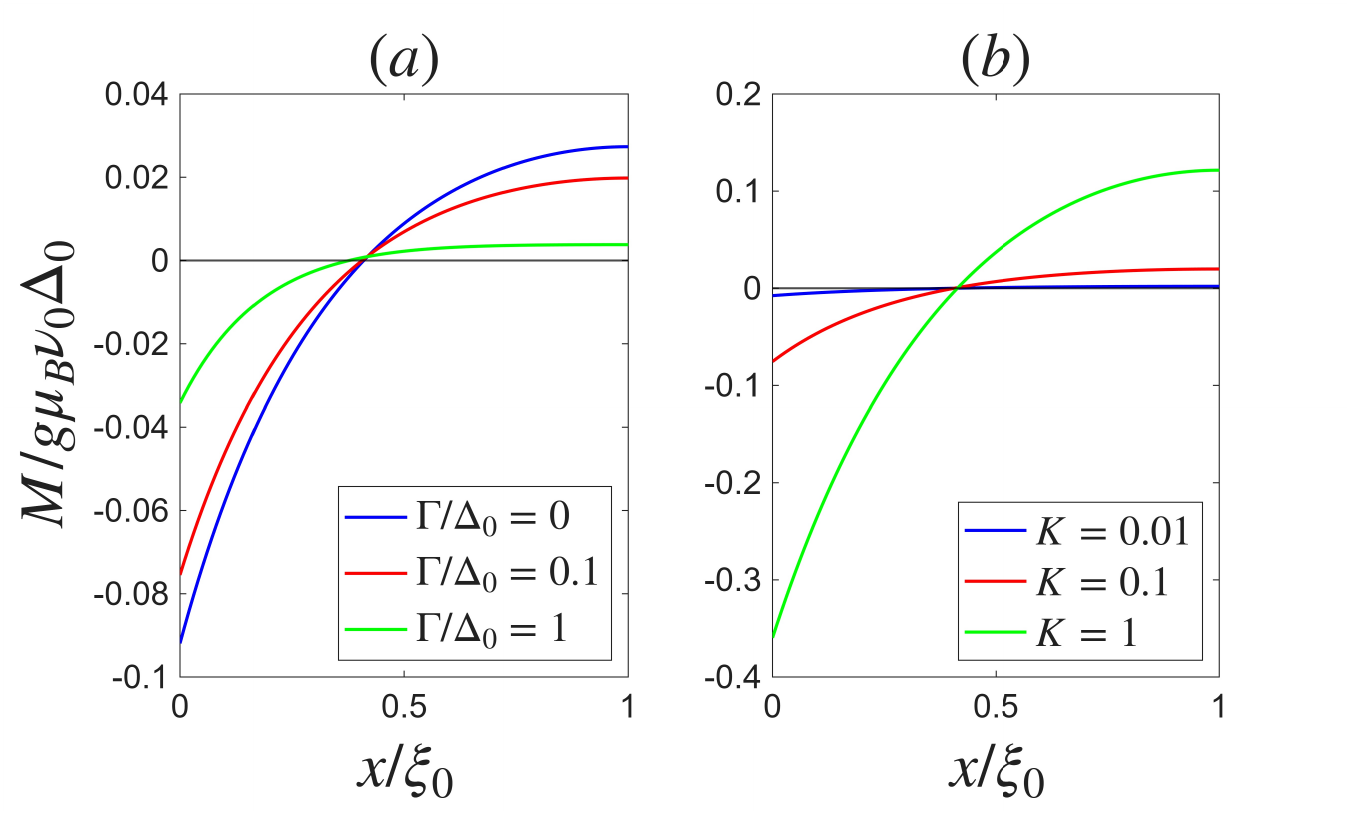}
    \caption{(a) Magnetization for $K = 0.1$ and different values of $\Gamma$. Increasing $\Gamma$ suppresses the overall magnitude of the magnetization and reduces the characteristic decay length $\xi$. (b) Magnetization for $\Gamma / \Delta_0 = 0.1$ and varying $K$. A larger $K$ enhances the induced magnetization. Around $x\approx\xi \approx 0.4\xi_0$, the magnetization changes sign, reflecting oscillations of the triplet component. The remaining parameters are $L = \xi_0$, $T / \Delta_0 = 0.1$, and $\gamma_B = 1$. The red line in both panels corresponds to the parameters of the low-energy model discussed in Sec.~\ref{Microscopicderiv}, with $h / \Delta_0 = 1$ and $\tau \Delta_0 = 0.1$.} 
    \label{SAM_Malpha0}
\end{figure}

\begin{figure}[H]
    \centering
    \includegraphics[width=0.9\linewidth]{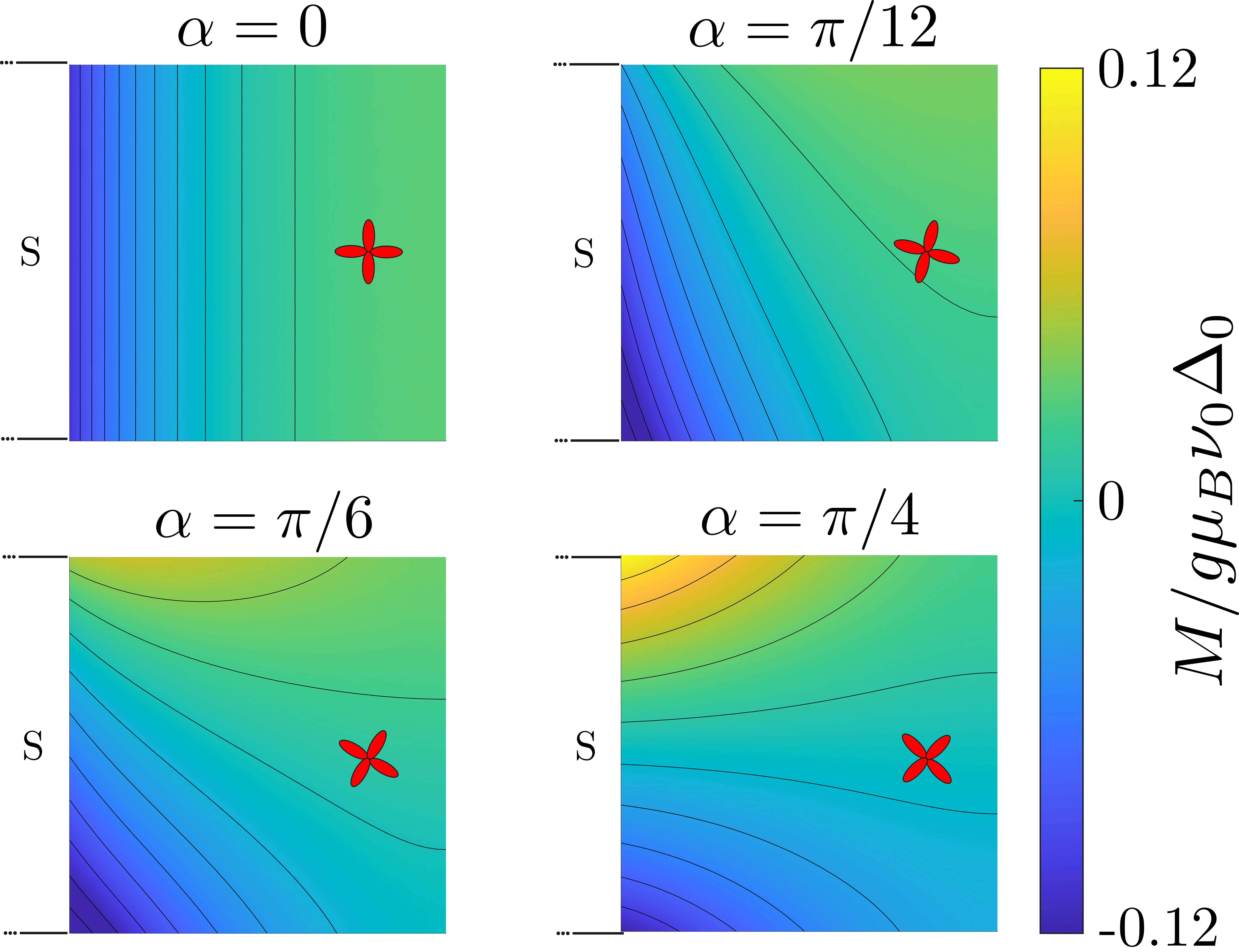}
    \caption{Magnetization in a 2D square altermagnet of side $L=\xi_0$ in contact with a superconductor for different values of $\alpha$. For $\alpha=0$, the induced magnetization is uniform along the y-direction and decays
    exponentially away from the interface. As $\alpha$ increases, a spatial redistribution of the magnetization emerges: negative
    magnetization localizes near the bottom-left corner, while positive magnetization develops near the top-left corner. At
    $\alpha=\pi/4$, d-wave symmetry dictates that the magnetization has equal magnitude and opposite sign in these two corners. For the other parameters we have chosen: $T/\Delta_0=0.1$, $\gamma_B=1$, $K=0.1$ and $\Gamma/\Delta_0=0.1$, which corresponds to $h/\Delta_0=1$ and $\tau\Delta_0=0.1$ in the low-energy model of Sec. \ref{Microscopicderiv}.} 
    \label{fig:SAM_alpha}
\end{figure}

The characteristic decay length of the condensate is given by $\xi_n=\operatorname{Re}(1/\kappa_{\omega_n}^\pm)$. If  $K\ll1$, it reduces to $\xi_n\approx\sqrt{D/(2(|\omega_n|+\Gamma))}$ coinciding with the behavior expected for a conventional antiferromagnet \cite{fyhn2023quasiclassical,TimNLSM}. 

\newpage

The parameter $K$ introduces an imaginary component to the pair amplitudes in Eq.~\eqref{pairs}, thereby generating a finite triplet component of the condensate that decays over the length scale $\xi_n$. Its generation can be traced back to Eq.~\eqref{MagnetizationGL}, where a spatial gradient of the order parameter induces a magnetization via the tensor $K_{jk}$. 

Consequently, these triplet correlations produce a magnetization along the $z$ axis, given by \cite{TimNLSM}:
\begin{align}
\label{magnetization}
 M&=\frac{1}{2}\nu_0 g\mu _B i\pi T\sum_{n=-\infty}^\infty \operatorname{Tr}(\check\tau_3\hat\sigma_3\check g_{w_n})\notag\\
    &\simeq\nu_0 g\mu_Bi\pi T\sum_{n=0}^{\infty}(f_-\tilde f_--\tilde f_+f_+)+O(f_+^2\tilde f_+^2-\tilde f_-^2f_-^2)\ .
\end{align}

Substituting the pair amplitudes from Eq.~\eqref{pairs} yields:
\begin{widetext}
\begin{align}\label{SAM_magnet}
    M_{\alpha=0}&=\frac{g\mu_B\nu_0\pi TD|\Delta|^2}{\gamma_B^2\xi_0^2}\sum_{n=0}^\infty \operatorname{Im}\left(\frac{\cosh^2(\kappa_{\omega_n}^+(x-L))}{(\omega_n+\Gamma)(\omega_n^2+|\Delta|^2)(1+iK)\sinh^2(\kappa_{\omega_n}^+L)}\right),
\end{align}
\end{widetext}
where $\operatorname{Im(\cdot)}$ denotes the imaginary part. 

\begin{figure}[b]
    \centering
    \includegraphics[width=0.8\linewidth]{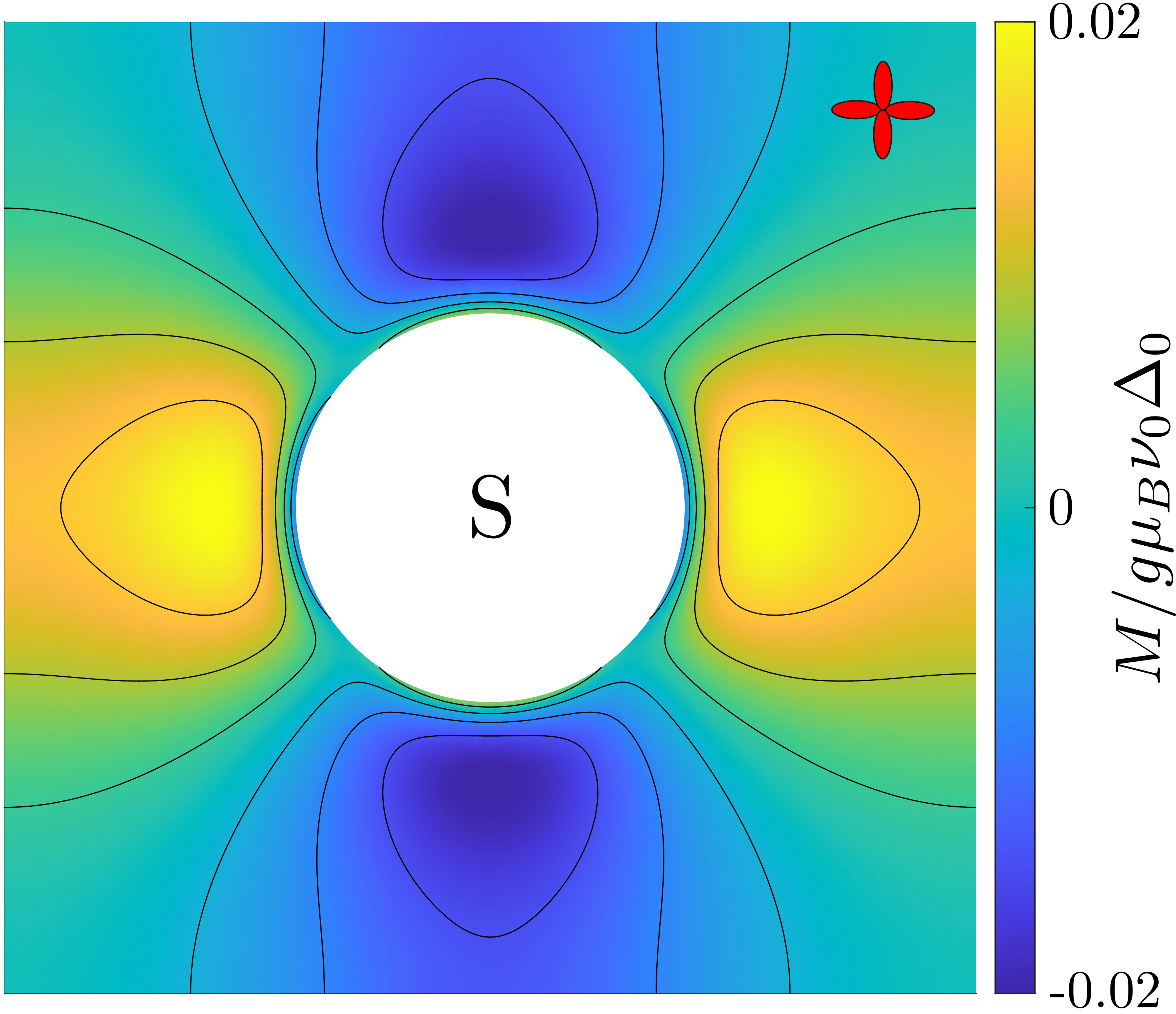}
    \caption{Circular superconducting island (white circle) of radius $r=\xi_0$ on top of a square altermagnet of side $L=5\xi_0$. The underlying $d$-wave symmetry of the altermagnet becomes particularly evident, giving rise to four distinct lobes in the magnetization with alternating sign. Magnetization oscillations are especially prominent in this geometry. For instance, along the horizontal direction, the magnetization at the interface with the superconductor is initially negative (blue) and then oscillates to a bigger positive (yellow) lobe. For the other parameters we have chosen: $T/\Delta_0=0.1$, $\gamma_B=1$, $K=0.1$ and $\Gamma/\Delta_0=0.1$, which corresponds to $h/\Delta_0=1$ and $\tau\Delta_0=0.1$ in the low-energy model discussed in Sec. \ref{Microscopicderiv}.} 
    \label{SAM_island}
\end{figure}

Thus, a finite magnetization emerges in the altermagnet, decaying exponentially over a distance $\xi_0$. To illustrate this behavior, in Fig.~\ref{SAM_Malpha0} we plot the magnetization, determine by Eq.~\eqref{SAM_magnet}, for different values of $K$ and $\Gamma$. As expected, the magnetization decreases monotonically with increasing $\Gamma$ or decreasing $K$. Moreover, the magnetization exhibits a sign change at approximately $x \approx \xi_0$, a feature that is remarkably robust and persists even for very small $K$. Beyond this first sign change, additional oscillations may occur, although they are typically strongly suppressed due to the exponential decay. According to Fig.~\ref{SAM_Malpha0}, the induced magnetic moment can reach values of the order of $g\mu_B\nu_0\Delta_0$ for sufficiently large $K$. Such a magnetic moment should be measurable by means of the polar Kerr effect, as used in Ref.~\cite{xia2009inverse} to detect the small magnetization induced in a superconductor via the inverse proximity effect \cite{bergeret2004induced}.

\begin{figure}[b]
    \centering
    \includegraphics[width=0.9\linewidth]{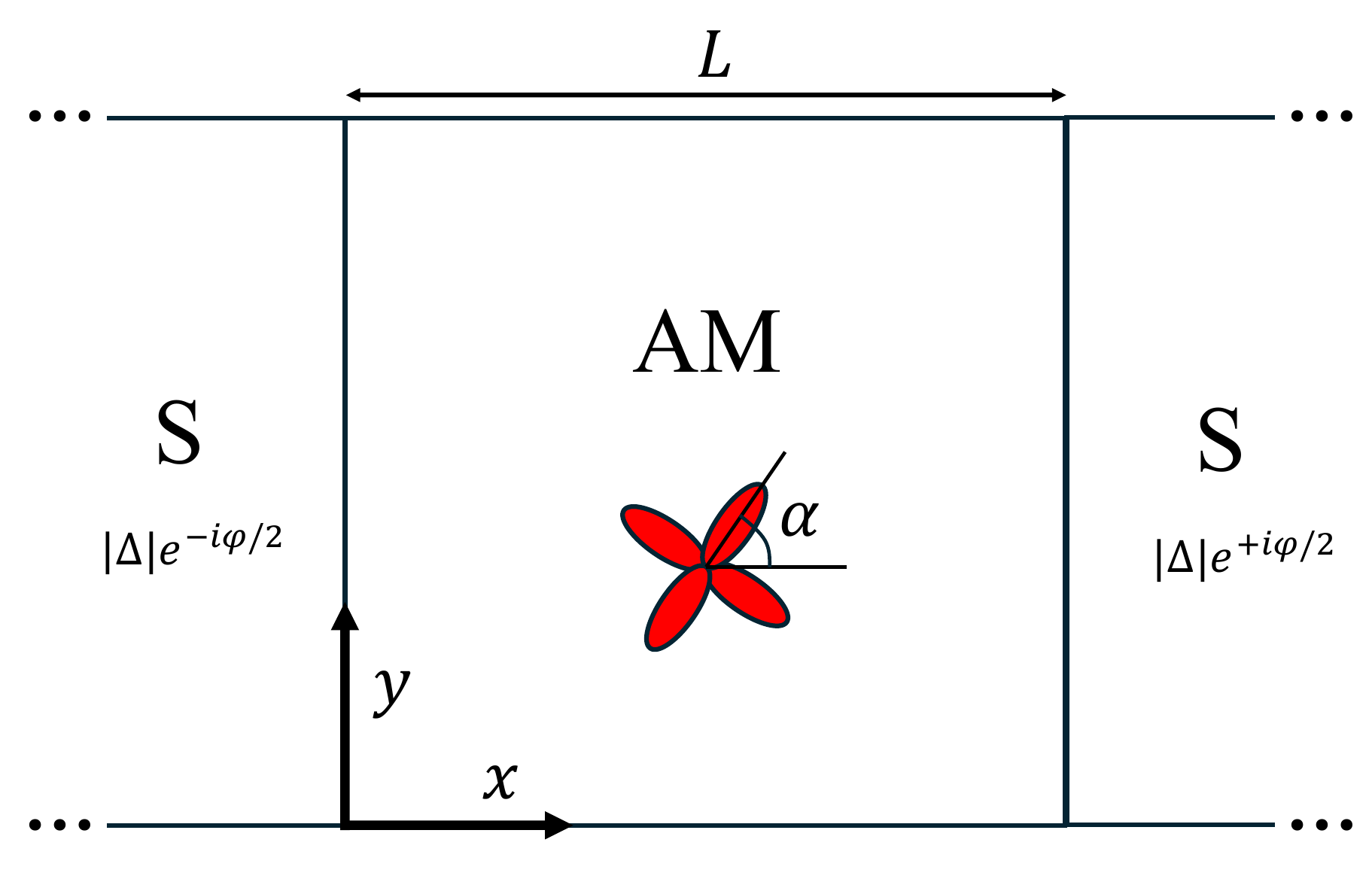}
    \caption{A square altermagnet of dimensions $[0,L]\times[0,L]$ sandwiched between two infinite (bulk) superconductors. At $x=0$ the superconducting order parameter is 
    $\check\Delta=|\Delta|e^{-i\check\tau_3\varphi/2}\check \tau_1$, while at $x=L$ it is 
    $\check\Delta=|\Delta|e^{i\check\tau_3\varphi/2}\check \tau_1$, such that the total phase difference across the junction is $\varphi$. The remaining boundaries at $y=0$ and $y=L$ are taken to be vacuum.} 
    \label{SAMS_sketch}
\end{figure} 

In Fig.~\ref{fig:SAM_alpha}, we present numerical results for the magnetization for four different altermagnet orientations. For $\alpha=0$ Eq.~\eqref{SAM_magnet} applies and the magnetization is uniform along $y$. As $\alpha$ increases, opposite-sign magnetization accumulates at the S/AM interface corners. At $\alpha=\pi/4$, the two corners carry equal but opposite sign magnetization, highlighting the underlying $d$-wave symmetry of the altermagnet.

The $d$-wave nature of the induced magnetization becomes even clearer if one considers a superconducting island of radius $r$ deposited on an altermagnetic substrate, see Fig.~\ref{SAM_island}. In this case, four lobes with alternating sign appear around the island, a remarkable manifestation of microscopic $d$-wave symmetry on the mesoscopic scale $\sim\xi_0$. In addition, the first sign change is more pronounced in this geometry.

\subsection{S/AM/S Josephson junction}\label{sec:junction}
We now consider a two-dimensional S/AM/S Josephson junction consisting of a square altermagnet of side length $L$ sandwiched between two bulk (infinite) superconductors with a total phase difference $\varphi$, as illustrated in Fig.~\ref{SAMS_sketch}.

As in the case of the S/AM bilayer, solving Eq.~\eqref{LinearUsadel} together with the boundary conditions for this case (see Appendix~\ref{KL_SAMS}) generally requires numerical methods.
However, for the orientation $\alpha=0$, the problem becomes independent of the coordinate $y$, allowing for analytic solutions. In this case, the induced magnetization reads:
\begin{widetext}
\begin{equation}\label{eq:SAMS_M}
    M_{\alpha=0} = \frac{g\mu_B\nu_0\pi TD|\Delta|^2}{\gamma_B^2\xi_0^2}\sum_{n=0}^\infty \operatorname{Im}\Bigg\{\frac{[\cosh(\kappa^+_{\omega_n}x)+\cosh(\kappa^+_{\omega_n}(x-L))]^2-2(1-\cos(\varphi))\cosh(\kappa^+_{\omega_n}x)\cosh(\kappa^+_{\omega_n}(x-L))}{(\omega_n+\Gamma)(\omega_n^2+|\Delta|^2
    )(1+iK) \sinh^2(\kappa_{\omega_n}^+L)}\Bigg\}\;.
\end{equation}
\end{widetext}

The expression inside the summation contains two distinct contributions. The first term in the curly brackets corresponds to the proximity induced magnetization (PIM) discussed for the S/AM bilayer, with contributions from both superconducting interfaces located at $x = 0$ and $x = L$. The second term, proportional to $1-\cos(\varphi)$, represents the nonlinear magnetoelectric effect \cite{zyuzin2024magnetoelectric}, in which a phase difference induces a magnetization. Importantly, the resulting magnetization is even in the superconducting phase difference, in contradiction to the claims made in Ref. \cite{giil2024quasiclassical}.  An equilibrium phase-odd magnetization is forbidden  by  symmetry \cite{TimNLSM,kokkeler2025nonequilibrium}. Consequently, from a symmetry perspective, this effect is distinct from the spin-splitting mechanisms that occur in the normal state. 

\begin{figure}[H]
    \centering
    \includegraphics[width=0.9\linewidth]{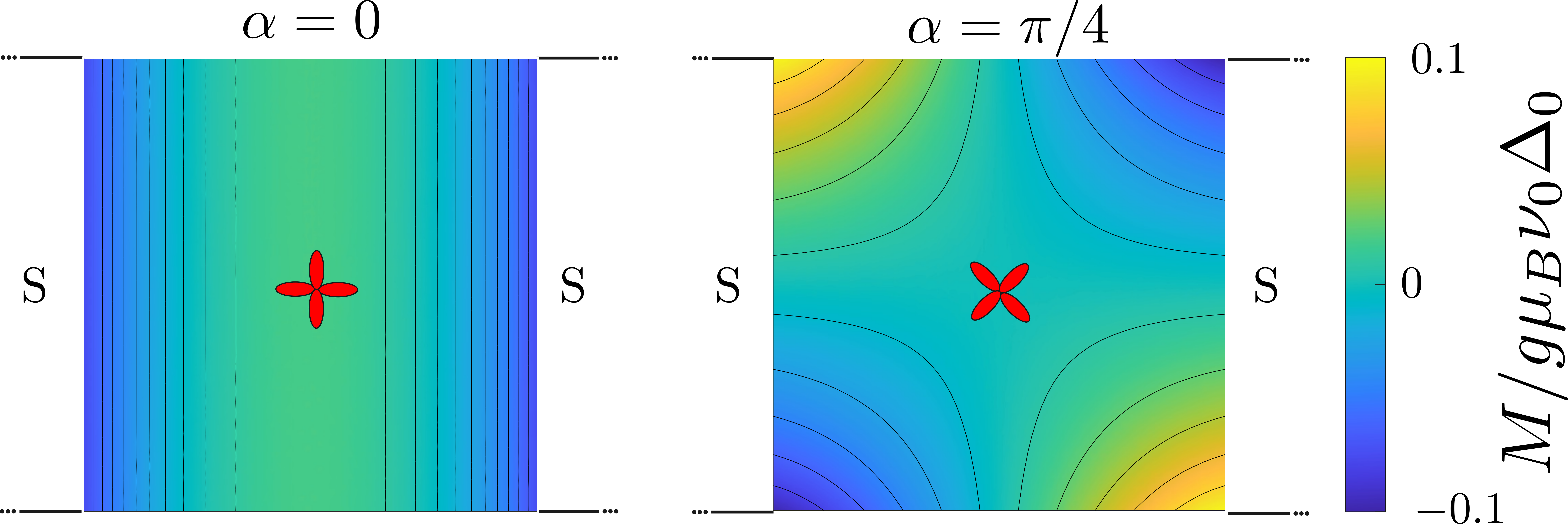}
    \caption{Magnetization in a square S/AM/S junction of side $L=\xi_0$. The phase difference between the two superconductors is $\varphi=\pi/2$. The magnetization near each interface resembles the bilayer case: it peaks at the interface and then decays exponentially. For $\alpha=0$ the magnetization is $y$-independent and for $\alpha=\pi/4$ the magnetization at all four corners reaches equal magnitude with alternating sign, consistent with the underlying $d$-wave symmetry. For the other parameters we have chosen: $T/\Delta_0=0.1$, $\gamma_B=1$, $K=0.1$ and $\Gamma/\Delta_0=0.1$, which corresponds to $h/\Delta_0=1$ and $\tau\Delta_0=0.1$ in the low-energy model of Sec. \ref{Microscopicderiv}.} 
\label{fig:SAMS_alpha}
\end{figure}

In Fig.~\ref{fig:SAMS_alpha}, we show numerical results for the magnetization at different crystallographic orientations. Near each interface the profile resembles the S/AM bilayer case from Fig.~\ref{fig:SAM_alpha}. For $\alpha=0$ the magnetization is uniform along $y$, while for $\alpha=\pi/4$ a four-lobe structure with alternating sign emerges, reflecting the $d$-wave symmetry.

For $\alpha=0$, one can also derive an analytic expression for the Josephson current:
\begin{widetext}
    \begin{equation}\label{supercurrent}
        j_x=\frac{\sigma_n}{2eD}i\pi T\sum_{n=0}^{\infty}\operatorname{Tr}(\check{\tau}_3\check{\mathcal{J}}_x)=\frac{\sigma_n\pi TD|\Delta|^2}{e\gamma_B^2\xi_0^2}\sin(\varphi)\sum_{n=0}^\infty\operatorname{Re}\left(\frac{\kappa_{\omega_n}^+}
        {(\omega_n+\Gamma)(\omega_n^2+|\Delta|^2)\sinh(\kappa_{\omega_n}^+L)}\right).
    \end{equation}
\end{widetext}

\begin{figure}[H]
    \centering
    \includegraphics[width=\linewidth]{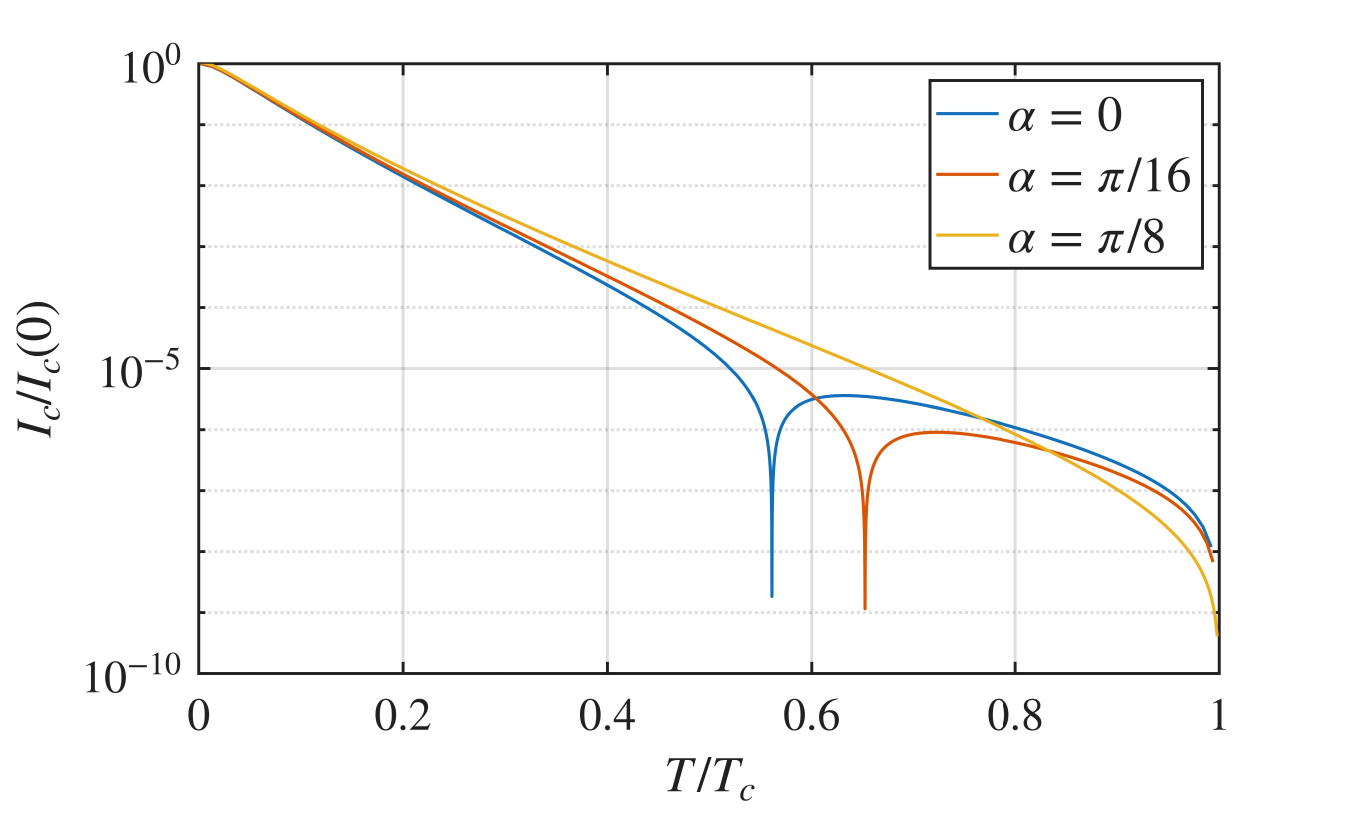}
    \caption{Critical current $I_c$ in a S/AM/S rectangular junction of size $L_x=10\xi_0$ and $L_y=\xi_0$ as a function of the temperature for different altermagnet orientations. A logarithmic scale was chosen for the vertical axis for clarity of presentation. For $\alpha=0$ there is a sharp dip corresponding to a $0-\pi$ transition at $T/T_c\approx0.56$ and for $\alpha=\pi/16$ at $T/T_c\approx0.65$. Notably, for $\alpha=\pi/8$, superconductivity is suppressed before the transition can occur. For the other parameters we have chosen: $\gamma_B=1$, $K=0.2$ and $\Gamma/\Delta_0=0.2$ (which corresponds to $\tau\Delta_0=0.2$ and $h/\Delta_0=1$ in the low-energy model of Sec. \ref{Microscopicderiv}). The dependence of the order parameter on temperature is approximately given by \cite{tinkham2004introduction}: $|\Delta|\approx\Delta_0\tanh\left(1.74\sqrt{T_c/T-1}\right)$ with $T_c\approx0.57\Delta_0$.} 
    \label{SAMS_zeropi}
\end{figure}

A notable feature of Eq.~\eqref{supercurrent} is the term $\sinh(\kappa_{\omega_n}^+L)$ in the denominator, which can change sign because $\kappa_{\omega_n}^+$ is a complex number for $K\neq0$. This sign reversal is indicative of $0-\pi$ transitions \cite{bulaevskii1977superconducting,ryazanov2001coupling}. Figure~\ref{SAMS_zeropi} illustrates this: the critical current $I_c$ as a function of temperature $T$ vanishes at a single point, consistent with a $0-\pi$ transition. The transition is most clearly visible for $\alpha=0$ and becomes progressively suppressed as $\alpha$ increases. For $\alpha=\pi/8$ no transition occurs for the parameters we have chosen.

The mechanism driving this $0$–$\pi$ transition is  similar to the well-known superconductor/ferromagnet/superconductor (S/F/S) case \cite{RevModPhys.77.935}: the exchange field in the F layer causes the pair amplitudes to oscillate in space, \textit{i.e.}, the corresponding inverse length $\kappa_{\omega_n}^{\pm}$ acquire an imaginary component. 
In the S/AM/S junction the effective exchange is anisotropic, it has  additional  dependence on $\alpha$. For  $\alpha=0$ the altermagnetic band splitting along the transport direction ($x$ axis) is maximized, thereby enhancing the mechanism responsible for the $0$–$\pi$ transition. In contrast, for $\alpha=\pi/4$ the altermagnetic bands are not spin-split along the $x$ direction and the $0$–$\pi$ transition is suppressed.

Previous studies have predicted $0-\pi$ transitions in altermagnets for clean systems \cite{ouassou2023dc,zhang2024finite,PhysRevB.108.075425,lu2024varphi,PhysRevB.111.165406}. However, to the best of our knowledge, here we show for the first time that this effect can persist even in the diffusive limit. 

\section{Low energy effective model}\label{Microscopicderiv}
So far, all the results presented are obtained from the Usadel equation Eq.~\eqref{UsadelAltermagnet}, which is valid for any altermagnet with disorder, independently of its microscopic details. What will change from one microscopic realization to another are the phenomenological constants $K_{ajk}$ and $\Gamma_{ab}$.

In this section, to make a connection with previous works, we discuss the relation between the parameters of our model and those of a specific low-energy effective Hamiltonian when disorder is taken into account. Specifically, in Appendix~\ref{microscopic} we derive the Usadel equation from a Bogoliubov--de Gennes (BdG) Hamiltonian describing a collinear $d$-wave altermagnet aligned along the $z$ axis. This BdG Hamiltonian has been used in many works \cite{lu2024varphi,hu2025nonlinear,zyuzin2024magnetoelectric,chourasia2025thermodynamic,sharma2025tunable} to describe ballistic altermagnets:
\begin{equation}\label{altermagnethamiltonian}
\check H_{\text{BdG}}(\mathbf{r})=\left[
-\frac{\nabla_{\mathbf{r}}^2}{2m}-\mu \right] \check\tau_3+ \frac{h_{ij}}{2 p_F^2} \hat\sigma_3\partial_{r_i} \partial_{r_j} -\check{\Delta} + \check V_{\text{imp}}\;, 
\end{equation}
where $h_{ij}$ is the anisotropic exchange field characterizing altermagnetism, and we have explicitly added the scattering potential $\check{V}_{\mathrm{imp}}$ due to nonmagnetic impurities.

In Appendix~\ref{microscopic}, we show that an Usadel equation of the form given in Eq.~\eqref{UsadelAltermagnet} can be derived from the above Hamiltonian in the diffusive limit. For this particular realization, the tensor coefficients are related to the parameters of the Hamiltonian as follows:
\begin{equation}\label{values}
K_{3jk}=\frac{6\tau}{d+2}h_{jk}\;,\ \Gamma_{33}=\frac{2h^2\tau}{d(d+2)}\;,\ T_{3jk}=\frac{h_{jk}}{2\mu}\;, 
\end{equation}
where $\tau$ is the elastic scattering time due to $V_{\text{imp}}$ within the Born approximation, $h^2=h_{ij}h_{ji}/2$, and $d$ is the dimensionality of the sample. All components of the tensors that correspond to a spin index other than $3$ vanish, consistent with previous symmetry arguments \cite{TimNLSM}. 

Notice that within this model,  the tensors $K_{ajk}$ and $\Gamma_{ab}$ scale linearly with $\tau$, analogous to the dependence of the anomalous Hall or spin–Hall conductivities on disorder in system with spin–orbit–coupling. Moreover, the tensor $T_{3jk}$, in contrast to $K_{ajk}$ and $\Gamma_{ab}$, is of next leading order in the quasiclassical limit, as it is proportional to $h/\mu$.

The spin–relaxation term $\Gamma_{33} \propto \tau$ describes a D’yakonov–Perel’–type mechanism~\cite{d1971spin}.
Substituting this form into the self-consistency relation Eq.~\eqref{Selfconsistency} shows that increasing disorder ({\it i.e.}, decreasing $\tau$) enhances the critical temperature $T_c$, consistent with the predictions of Ref.~\cite{vasiakin2025disorder}. 
It is worth emphasizing that the $\Gamma$ tensor in our effective action, Eq. (\ref{ActionAltermagnet}), is not bound to be related only to this mechanisms, but it may have other contributions, as for example those studied in Refs. \cite{fyhn2023quasiclassical,sedov2025quantum}.

Most of the results presented in Figs.~(\ref{SAM_Malpha0}--\ref{SAM_island}, \ref{fig:SAMS_alpha}--\ref{SAMS_zeropi}) 
are obtained for, $h/\Delta_0 = 1$, $\tau \Delta_0 = 0.1$, and low transparency interfaces, {\it i.e.} within the diffusive regime in this example. This demonstrates that the signatures of altermagnetism, such as the proximity induced magnetization, are also present in disordered systems. Naturally, in cleaner samples or with higher interface transparency, we expect these effects to become even more pronounced.

It should be noted that all the results above also apply qualitatively if, instead of an AM, one considers the proximity effect in a normal metal grown on top of an insulating AM. In the case of thin normal films, parameters such as the exchange field or the relaxation length take effective values that scale inversely with the film thickness, in analogy with superconductor–ferromagnetic insulator systems~\cite{heikkila2019thermal,bobkov2025inverse}.

\section{Conclusions}\label{sec:Conclusions}
In this article we investigate the interplay between superconductivity and altermagnetism, particularly in S/AM heterostructures. Starting from the Luttinger-Ward-type free energy functional, based on the nonlinear sigma model, we derive the corresponding GL free energy and identified two key contributions: a usual pair-breaking term stemming from a spin relaxation,  and a higher order gradient term. The latter produces two distinct nonlinear effects. On the one hand, a magnetoelectric effect that generates a finite magnetization from a phase gradient. This effect is quadratic in the phase gradient and has been studied in ballistic systems in Refs.~\cite{hu2025nonlinear,zyuzin2024magnetoelectric}. On the other hand, a magnetic moment can also be generated by inhomogeneities of the modulus of the order parameter. This latter effect explains the proximity-induced magnetization predicted in Sec.~\ref{S/AMhetero}.  

We also demonstrate that in S/AM heterostructures, superconducting correlations penetrating the altermagnet induce an equilibrium magnetization localized near the interface. This magnetization changes its sign  over a length comparable to the coherence length.  In S/AM/S Josephson junctions, both amplitude and phase gradients of $\Delta$ contribute to the magnetization, leading to the coexistence of the two effects. The Josephson current may also exhibit $0$-$\pi$ transitions, even in the diffusive regime.  

In summary, our theory provides a foundation for studying the equilibrium and transport properties of disordered systems that combine superconductivity and altermagnetic order.

\section*{Acknowledgments}
R.~H. and F.~S.~B. thank financial support from the the Spanish MCIN/AEI/10.13039/501100011033 
through the grants PID2023-148225NB-C31, and TED2021-130292B-C41. F. S. B. also thanks financial support from the
European Union’s Horizon Europe research and innovation program under grant agreement No. 101130224 (JOSEPHINE). I.~V.~T. acknowledges support from the
Spanish MCIN/AEI/10.13039/501100011033 through
the Project No. PID2023-148225NB-C32, and the
Basque Government (Grant No. IT1453-22). T. K. acknowledges support by the Research Council of Finland through DYNCOR, Project Number 354735 and through the Finnish Quantum Flagship, Project Number 359240. S.~I. is supported by the Research Council of Finland (Grant No. 355056). 
The work by T.K. and S. I. is part of the Finnish Centre of Excellence in Quantum Materials (QMAT, Project No. 374165).

\bibliography{biblio} 

\onecolumngrid
\appendix

\section{Microscopic derivation of the Usadel equation}\label{microscopic}
In this Appendix, we derive the Usadel equation for a superconductor with altermagnetic order. We will arrive at the result directly from the Gor'kov-Dyson equation, in contrast to the usual approach \cite{belzig1999quasiclassical} where one first derives the Eilenberger equation and then obtains the Usadel equation by taking the diffusive limit. Our approach, although somewhat more complicated than the standard route, allows us to derive systematically and in a controlled way also the terms beyond the leading order in the quasiclassical approximation.

\subsection{Gor'kov-Dyson equation}
We consider a superconductor/altermagnet structure described by the following BdG Hamiltonian in the real space
\begin{equation}
\check H_{\text{BdG}}(\mathbf{r})=\left[
-\frac{\partial_{\mathbf{r}}^2}{2m}-\mu \right] \check\tau_3+ \frac{h_{ij}}{2 p_F^2} \hat\sigma_3\partial_{r_i} \partial_{r_j} -\check{\Delta}+\check{V}_{\text{imp}}. 
\end{equation}
For simplicity, we focus on the case $h_{xy}=h_{yx}=h \neq 0$, with $h_{ij}=0$ otherwise.
The Green's function $\check G$ describing our system in the presence of disorder is determined from the Gor'kov-Dyson equation (GDE):
\begin{equation}
(i \omega_n -\check H_{\text{BdG}}(\mathbf{r})-\check \Sigma_{\text{imp}}(\mathbf{r})) \check G(\mathbf{r},\mathbf{r'})=\delta(\mathbf{r}-\mathbf{r'}),
\end{equation}
where $\check \Sigma_{\text{imp}}$ is the impurity self-energy. To proceed, we go to the Wigner representation,  $\check G(\mathbf{r},\mathbf{r'})=\sum_{\mathbf{p}} e^{-i \mathbf{p} (\mathbf{r-r'})} \check G_\mathbf{p}\left(\frac{\mathbf{r+r'}}{2}\right)$, in which the GDE becomes:
\begin{equation}\label{eq:rewrite}
\left(-\xi(\mathbf{p})+\check \Omega(\mathbf{R}) +\check{H}  \frac{p_xp_y}{p_F^2} +\frac{i}{2\tau} \check \tau_3 \bar{G}(\mathbf{R})\right) e^{\frac{i}{2} (\overleftarrow{\partial}_\mathbf{R} \overrightarrow{\partial}_{\mathbf{p}}-\overleftarrow{\partial}_\mathbf{p} \overrightarrow{\partial}_{\mathbf{R}} )} \check \tau_3 \check G_\mathbf{p}(\mathbf{R})=1.
\end{equation}
Here, we introduced the center of mass coordinate $\mathbf{R}=\frac{\mathbf{r}+\mathbf{r'}}{2}$, $\check \Omega (\mathbf{R})=(i\omega_n+\check{\Delta} (\mathbf{R})) \check \tau_3$, $\check{H}=h\hat\sigma_3 \check\tau_3$ and $\xi(\mathbf{p})=\frac{\mathbf{p}^2}{2m}-\mu$. The last term in the parentheses in Eq.~\eqref{eq:rewrite} comes from the impurity self-energy in the self-consistent Born approximation, where the "bar" operation denotes integration over momenta $\bar{G}(\mathbf{R})=\frac{i }{\pi \nu_0} \sum_\mathbf{p} \check G_\mathbf{p}(\mathbf{R})$. The right (left) arrows on top of the partial derivative operators mean that they only act to the right (left). Next, using the fact that the action of gradient operators in the exponent can be treated as translations, namely $A(\mathbf{p,R}) e^{\frac{i}{2} (\overleftarrow{\partial}_\mathbf{R} \overrightarrow{\partial}_{\mathbf{p}}-\overleftarrow{\partial}_\mathbf{p} \overrightarrow{\partial}_{\mathbf{R}} )} B(\mathbf{p,\mathbf{R}})=A(\mathbf{p}-\frac{i}{2}\partial_\mathbf{R},\mathbf{R}+\frac{i}{2}\partial_\mathbf{p})B(\mathbf{p},\mathbf{R})$, Eq.~\eqref{eq:rewrite} becomes:
\begin{equation}
\left[
-\xi({\mathbf{p}-\frac{i}{2}\partial_{\mathbf{R}}})+ \frac{\check H}{p_F^2} (p_x-\frac{i}{2}\partial_{R_x})(p_y-\frac{i}{2}\partial_{R_y})+ \check \Omega(\mathbf{R}+\frac{i}{2} \partial_\mathbf{p})+\frac{i}{2\tau}\check\tau_3 \bar{G}(\mathbf{R}+\frac{i}{2}\partial_\mathbf{p})
\right] \check\tau_3 \check G_\mathbf{p}(\mathbf{R})=1.
\end{equation}

To simplify upcoming calculations, it is convenient to introduce dimensionless parameters $\eta=2\tau \xi(\mathbf{p})$ and $\psi=1/(2\tau \mu)$, and the dimensionless Green's function $\check{\mathcal{G}}_\mathbf{p}(\mathbf{R})=\frac{1}{2\tau}\check\tau_3 \check G_\mathbf{p}(\mathbf{R})$. Then, we obtain the dimensionless GDE:
\begin{multline}
\bigg[
-\eta +i l \sqrt{1+\eta\psi} n_i \partial_{R_i}+\frac{\psi l^2}{4}\partial^2_{\mathbf{R}} + 2 \tau\check{H} (n_x \sqrt{1+\eta\psi}-\frac{i\psi l}{2} \partial_{R_x}) (n_y \sqrt{1+\eta\psi}-\frac{i\psi l}{2} \partial_{R_y}) \\
+ 2 \tau \check{\Omega}(\mathbf{R}+\frac{i}{2}\partial_\mathbf{p})+ i  \bar{\mathcal{G}}(\mathbf{R}+\frac{i}{2}\partial_\mathbf{p}) 
\bigg] \check{\mathcal{G}}_\mathbf{p}(\mathbf{R})=1,
\end{multline}
with $l=v_F\tau$, $\mathbf{n}=\mathbf{p}/|\mathbf{p}|$, and $p_i=\sqrt{2m(\mu+\xi)}n_i=p_F\sqrt{1+\eta\psi}n_i$. The momentum averaging operation can now be represented as:
\begin{equation}
\bar{\mathcal{G}}=\frac{i}{\pi \nu_0} \int_{-1/\psi}^{\infty} d\eta \, \nu(\eta) \langle \mathcal{G} \rangle, 
\label{Eq:avg}
\end{equation}
where the density of states is $\nu(\eta)=\nu_0 (1+\eta\psi)^{d/2-1}$, $d$ is the dimension of the system, and $\langle...\rangle$ stands for averaging over the angles $n_i$. The momentum derivative can be written as $\partial_{p_i}=n_i \partial_p+\frac{1}{p} (\partial_{n_i}-n_i n_j \partial_{n_j})$, or, in dimensionless variables,  $\partial_{p_i}=l \left( 2 n_i \sqrt{1+\eta\psi} \partial_\eta +\psi \frac{\partial_{n_i}-n_i n_j \partial_{n_j}}{\sqrt{1+\eta\psi}} \right)$.

\subsection{Gradient expansion and diffusive limit}
We next assume the quasiclassical approximation and the diffusive limit, meaning that the following hierarchy of energy scales holds $\mu\gg \tau^{-1}\gg |\Delta|, h$. Within these assumptions, the dimensionless parameter $\psi$ is small. In the quasiclassical theory one usually keeps only the leading order in $\psi$ $(\psi^0)$ \cite{belzig1999quasiclassical}, but here,  for completeness, we will also retain terms of the next order ($\psi^1$). 

The quasiclassical approximation allows us to do the gradient expansion. We will retain terms up to the second order in spatial gradients $\partial_{R_i}$, namely $\check{\mathcal{G}}(\mathbf{R}+\frac{i}{2}\partial_\mathbf{p})\approx \check{\mathcal{G}}+\frac{i}{2} \check{\mathcal{G}}_i \partial_{p_i}-\frac{1}{8} \check{\mathcal{G}}_{ij} \partial_{p_i} \partial_{p_j}$, 
and similarly for $\check \Omega(\mathbf{R+\frac{i}{2}\partial_\mathbf{p}})$. Here we use the notation $\check{\mathcal{G}}_i=\partial_{R_i} \check{\mathcal{G}}$, $\check{\mathcal{G}}_{ij}=\partial_{R_i} \partial_{R_j} \check{\mathcal{G}}$. Then, after some rearranging, the GDE can be rewritten as:
\begin{multline}
(-\eta+i \bar{\mathcal{G}})\check{\mathcal{G}}=1 - \bigg[
2\check{\Omega} \tau \check{\mathcal{G}}+\frac{i}{2} (2\tau \check{\Omega}_i+i \bar{\mathcal{G}}_i) \partial_{p_i}\check{\mathcal{G}}
-\frac{1}{8} (2\tau \check{\Omega}_{ij}+i \bar{\mathcal{G}}_{ij})\partial_{p_i}\partial_{p_j}\check{\mathcal{G}}+i l \sqrt{1+\eta\psi}n_i \check{\mathcal{G}}_i+\frac{\psi l^2}{4} \check{\mathcal{G}}_{ii}\\
+2\tau \check{H} n_x n_y (1+\eta\psi)\check{\mathcal{G}}-i\psi l \tau \check{H} \sqrt{1+\eta\psi} (n_x \check{\mathcal{G}}_y+n_y \check{\mathcal{G}}_x) 
\bigg].
\label{Eq:gradexp}
\end{multline}

Next, we expand the Green's function in  parameters $l \partial_R$, $\tau\check{\Omega}$ and $\tau \check{H}$, all of which are small in the diffusive limit:
\begin{equation}
\check{\mathcal{G}}=\check{\mathcal{G}}^{(0)}+\check{\mathcal{G}}^{(1)}+\check{\mathcal{G}}^{(2)}+\check{\mathcal{G}}^{(3)}+...,
\label{Eq:expansion}
\end{equation}
where $n$ in $\check{\mathcal{G}}^{(n)}$ denotes the order in small parameters. From Eq.~\eqref{Eq:gradexp}, we see that the zeroth-order contribution is given as:
\begin{equation}
\check{\mathcal{G}}^{(0)}=(-\eta+i\bar{\mathcal{G}}^{(0)})^{-1}.
\label{Eq:G0}
\end{equation}
Similarly, we can write equations for higher-order components. For $k\geq 1$, we have:
\begin{equation}
\check{\mathcal{G}}^{(k)}+i\check{\mathcal{G}}^{(0)} \bar{\mathcal{G}}^{(k)}\check{\mathcal{G}}^{(0)}=X^{(k)},
\label{Eq:Gk}
\end{equation}
with:
\begin{multline}
X^{(k)}=-\check{\mathcal{G}}^{(0)} \bigg[-i \theta_{k-s-1} \bar{\mathcal{G}}^{(s)} \check{\mathcal{G}}^{(k-s)}+
2\tau\check{\Omega} \check{\mathcal{G}}^{(k-1)}+i \theta_{k-2} \tau \check{\Omega}_i \partial_{p_i}\check{\mathcal{G}}^{(k-2)}-\frac{\theta_{k-s-1}}{2} \bar{\mathcal{G}}^{(s)}\partial_{p_i} \check{\mathcal{G}}^{(k-s-1)} \\- \frac{\theta_{k-3}}{4} \tau \check{\Omega}_{ij} \partial_{p_i}\partial_{p_j} \check{\mathcal{G}}^{(k-3)}
-\frac{i \theta_{k-s-2}}{8}\bar{\mathcal{G}}_{ij}^{(s)} \partial_{p_i}\partial_{p_j} \check{\mathcal{G}}^{(k-s-2)}+i l \sqrt{1+\eta \psi} n_i \check{\mathcal{G}}^{(k-1)}+
\frac{\theta_{k-2}\psi l^2 }{4} \check{\mathcal{G}}_{ii}^{(k-2)}
\\+2\tau \check{H}n_xn_y (1+\eta\psi)\check{\mathcal{G}}^{(k-1)}
-i\theta_{k-2}\psi l \tau \check{H}\sqrt{1+\eta \psi} \left(n_x \check{\mathcal{G}}_y^{(k-2)}+n_y \check{\mathcal{G}}_x^{(k-2)}\right)
\bigg],
\label{Eq:Xk}
\end{multline}
where we introduced the discrete step function $\theta_i=\begin{cases}1,\, i\geq0, \\ 0, \, i<0.\end{cases}$, and $s\geq 0$ is an internal summation variable (summation over repeated indices is assumed). 

\subsection{Solution procedure}
To solve Eq.~\eqref{Eq:gradexp}, we will exploit the fact that the momentum-averaged Green's function  $\bar{\mathcal{G}}$ of a diffusive system can be always written in the form $\bar{\mathcal{G}}=Q+B$, so that $Q^2=1$ and $[B,Q]=0$ \cite{virtanen2022nonlinear}. Namely, $Q$ is the part of the Green's function that is normalized to $1$, and $B$ is known as the longitudinal correction. We can expand $B$ up to third order in the small diffusive parameters, in the same way as in Eq.~\eqref{Eq:expansion}: $B\approx B^{(0)}+B^{(1)}+B^{(2)}+B^{(3)}$.

For clarity, in the following we will focus on 3D systems. Let us start by calculating $\check{\mathcal{G}}^{(0)}$. We first assume that $B^{(0)}$ vanishes, which will be justified \emph{a posteriori}. Then, from Eq.~\eqref{Eq:G0} we find $\check{\mathcal{G}}^{(0)}=(-\eta+i Q)^{-1}=-(\eta+i Q)/(\eta^2+1).$ We can now calculate $\bar{\mathcal{G}}^{(0)}$ by placing this expression into Eq.~\eqref{Eq:avg}. The resulting integral, as well as all other energy integrals we will encounter in the upcoming calculations, can be solved using residues \cite{virtanen2022nonlinear}:
\begin{equation}
\int_{-1/\psi}^{\infty} (1+\eta \psi)^{d/2-1} \frac{(1+\eta\psi)^m \eta^{n}}{(1+\eta^2)^p}= 
2\pi i \sum_{\pm} \text{Res}_{\eta=\pm i} q(\eta) \frac{(1+\eta\psi)^m \eta^{n}}{(1+\eta^2)^p}.
\end{equation}
The above expression is valid for all convergent integrals ($d,m,n,p$ are non-negative integers), and we introduced the function
\begin{equation}
q(\eta)=\begin{cases}
\frac{i}{2\pi} \ln (-1-\eta\psi)(1+\eta\psi)^{d/2-1}, d\,  \text{even} \\
\frac{i}{2}\sqrt{-1-\eta\psi}(1+\eta\psi)^{d/2-3/2}, d\, \text{odd}.
\end{cases}
\end{equation}
Note that the $Q$-independent part of $\check{\mathcal{G}}^{(0)}$ gives a divergence upon integration, but this divergence can be regularized by gauge-invariance arguments \cite{virtanen2022nonlinear}. Finally, expanding the result of the integration in $\psi$, we get $\bar{\mathcal{G}}^{(0)}=Q+O(\psi^2)$. This justifies our earlier assumption that $B^{(0)}=0$. From here, we also see that $B^{(i)}=\bar{\mathcal{G}}^{(i)}$ ($i=1,2,3...$).

We proceed to calculate the higher-order components, $\check{\mathcal{G}}^{(i)}$, $i=1,2,3$. From Eqs.~\eqref{Eq:Gk} and \eqref{Eq:Xk}, we have $\check{\mathcal{G}}^{(i)}+i  \check{\mathcal{G}}^{(0)} B^{(i)} \check{\mathcal{G}}^{(0)} =X^{(i)}$. We then apply the momentum averaging operation \eqref{Eq:avg} to this expression, and expand up to first order in $\psi$, to get $B^{(i)}+\frac{1}{2} (Q B^{(i)} Q-B^{(i)})-\frac{i\psi}{4} \{ Q,B^{(i)}\}=\bar{X}^{(i)}$. Then, using $Q^2=1$ and $[B^{(i)},Q]=0$, this is simplified to:
\begin{equation}
B^{(i)}=\bar{X}^{(i)}+\frac{i\psi}{4} (Q \bar{X}^{(i)}+\bar{X}^{(i)}Q),
\end{equation}
which is valid up to order $\psi^1$. Finally, we can express $\check{\mathcal{G}}^{(i)}$ as:
\begin{equation}
\check{\mathcal{G}}^{(i)}=X^{(i)}-i \check{\mathcal{G}}^{(0)} B^{(i)} \check{\mathcal{G}}^{(0)}.
\end{equation}

The Usadel equation is obtained from the condition that the longitudinal part commutes with $Q$, namely:
\begin{equation}
\frac{1}{2\tau}\sum_i [B^{(i)},Q]=0,
\end{equation}
where the prefactor $1/2\tau$ is added to get the standard form of the Usadel equation.

\subsection{Final solution and the Usadel equation}
The procedure described above is straightforward, but it produces a large number of terms, which we handle using symbolic computation software, \emph{Wolfram Mathematica} \cite{Mathematica}. In the following, we write the final result of this calculation. In the first order in small expansion parameters we get:
\begin{equation}
\frac{1}{2\tau}[B^{(1)},Q]=[Q,i \check{\Omega}].
\end{equation}
This is the standard commutator contribution to the Usadel equation. Similarly, in the second order in small parameters we have:
\begin{equation}
\frac{1}{2\tau}[B^{(2)},Q]=-D \partial_{R_i} (Q\partial_{R_i}Q)-\frac{\Gamma}{2} [Q,\hat \sigma_3\check \tau_3Q \hat \sigma_3\check \tau_3],
\end{equation}
where $D=\frac{1}{d} v_F^2\tau$ is the diffusion constant, and $\Gamma=\frac{2h^2\tau}{d(d+2)}$ is the relaxation rate. Here we can recognize the standard current term, and the D'yakonov-Perel-like spin-relaxation term. In the third order in the expansion we find
\begin{multline}
\frac{1}{2\tau}[B^{(3)},Q]= -\frac{3iD\tau}{2(d+2)}h_{ij} \partial_{R_i}[\hat\sigma_3 \check\tau_3+ Q\hat\sigma_3 \check\tau_3 Q, \partial_{R_j}Q]+\frac{D\tau \psi}{4} h_{ij} \partial_{R_i}\{\hat\sigma_3\check\tau_3+ Q \hat\sigma_3\check\tau_3 Q, Q \partial_{R_j}Q \} \\
-\frac{3i D \tau}{2(d+2)} h_{ij} [\hat\sigma_3\check\tau_3, Q\partial_{R_i} Q \partial_{R_j} Q]-\frac{D\tau \psi}{4} h_{ij} [\hat\sigma_3\check\tau_3, \partial_{R_i} Q \partial_{R_j} Q].
\end{multline}
In writing the above equations, we have kept only the leading-order contribution proportional to $\check{\Omega}$ that appears in $B^{(1)}$. The $\check{\Omega}$-dependent terms that appear in $B^{(2)}$ and $B^{(3)}$ were omitted for simplicity, since they are not relevant for the phenomena discussed here. 

Collecting the above results yields the final Usadel equation, $\frac{1}{2\tau}[B^{(1)}+B^{(2)}+B^{(3)},Q]=0$, which has a form specified in Eq.~\eqref{UsadelAltermagnet} with the coefficients given in Eq.~\eqref{values}. The generalization from the case discussed in this Appendix with only $h_{xy}=h_{yx}=h\neq0$ to Eq.~\eqref{values}, valid for any orientation, follows from rotational invariance of the tensor $h_{ij}$.

{
We note that the coefficients appearing in the third-order gradient contribution scale as $h_{ij} D \tau$ and $h_{ij} D \tau\psi$, respectively. This disagrees with an earlier derivation of the Usadel equation for altermagnets in the weak–proximity limit, $Q\approx \check\tau_3 + \begin{bmatrix}
    0&\hat f\\\hat{\tilde{f}}&0
\end{bmatrix}$, reported in Ref.~\cite{giil2024quasiclassical}, where the corresponding coefficients were found to scale instead as $h_{ij}D\tau\psi^2 \ll h_{ij} D \tau$ and $h_{ij} D \tau \psi$. The origin of this discrepancy lies in the truncation of the spherical-harmonic expansion of $\hat f$ in Ref.~\cite{giil2024quasiclassical}. Only the first angular harmonic was retained, whereas the second harmonic is also required and couples directly to the $d$-wave (second spherical harmonic) altermagnetic field. As a consequence, Ref.~\cite{giil2024quasiclassical} omitted a leading-order contribution to one of the coefficients.}

\section{Linearized Boundary Conditions}\label{Ap:BoundaryCons}
In Section \ref{S/AMhetero} of the main text, we solve the linearized Usadel equation, Eq.~\eqref{LinearUsadel} in a rectangular geometry where $(x,y)\in[0,L]\times[0,L]$. In this Appendix we present the  boundary conditions that complement the linearized equations which originate by substituting the parametrization  Eq.~\eqref{parametrization} into the Kupriyanov–Lukichev boundary condition Eq.~\eqref{eq:KL-BC} and retaining only terms linear in the pair amplitudes. Like in Sec.~\ref{S/AMhetero} in the main text, we parameterize the altermagnet tensor $K_{ajk}$ by a single magnitude $K$ and angle $\alpha$ as in Eq.~(\ref{eq:Kparameterization}). For the S/AM case (setup shown in Fig.~\ref{SAM_sketch} of the main text), to linear order in $1/\gamma_B$, we obtain:
\begin{subequations}\label{KL_SAM}
\begin{align}
    (1\pm i\operatorname{sgn}(\omega_n)K\cos(2\alpha))\partial_xf_{\pm}\pm i\operatorname{sgn}(\omega_n)K\sin(2\alpha)\partial_yf_{\pm}=-\frac{|\Delta|}{\gamma_B\xi_0\sqrt{\omega_n^2+|\Delta|^2}},\ & x=0
\\
   (1\pm i\operatorname{sgn}(\omega_n)K\cos(2\alpha))\partial_xf_{\pm}\pm i\operatorname{sgn}(\omega_n)K\sin(2\alpha)\partial_yf_{\pm}=0,\;\qquad \qquad \qquad \ \ & x=L
\\
    (1\mp i\operatorname{sgn}(\omega_n)K\cos(2\alpha))\partial_yf_{\pm}\pm i\operatorname{sgn}(\omega_n)K\sin(2\alpha)\partial_xf_{\pm}=0,\;\qquad \qquad \qquad \ \ & y=0,\, L
\end{align}
\end{subequations}
with analogous boundary conditions for $\tilde f_\pm$.

For the S/AM/S Josephson junctions (see Fig.~\ref{SAMS_sketch} of the main text), to linear order in $1/\gamma_B$, we obtain:
\begin{subequations}\label{KL_SAMS}
\begin{align}
    (1\pm i\operatorname{sgn}(\omega_n)K\cos(2\alpha))\partial_xf_\pm\pm i\operatorname{sgn}(\omega_n)K\sin(2\alpha)\partial_yf_\pm&=-\frac{|\Delta| e^{i\varphi/2}}{\gamma_B\xi_0\sqrt{\omega_n^2+|\Delta|^2}},\ x=0
\\
    (1\pm i\operatorname{sgn}(\omega_n)K\cos(2\alpha))\partial_xf_\pm\pm i\operatorname{sgn}(\omega_n)K\sin(2\alpha)\partial_yf_\pm&=\frac{|\Delta |e^{-i\varphi/2}}{\gamma_B\xi_0\sqrt{\omega_n^2+|\Delta|^2}},\quad x=L
\\
    (1\mp i\operatorname{sgn}(\omega_n)K\cos(2\alpha))\partial_yf_\pm\pm i\operatorname{sgn}(\omega_n)K\sin(2\alpha)\partial_xf_\pm&=0,\qquad \qquad \qquad \ \ \, y=0,\ L
\\
(1\pm i\operatorname{sgn}(\omega_n)K\cos(2\alpha))\partial_x\tilde{f}_\pm\pm i\operatorname{sgn}(\omega_n)K\sin(2\alpha)\partial_y\tilde{f}_\pm&=-\frac{|\Delta| e^{-i\varphi/2}}{\gamma_B\xi_0\sqrt{\omega_n^2+|\Delta|^2}},\ x=0
\\
(1\pm i\operatorname{sgn}(\omega_n)K\cos(2\alpha))\partial_x\tilde{f}_\pm\pm i\operatorname{sgn}(\omega_n)K\sin(2\alpha)\partial_y\tilde{f}_\pm&=\frac{|\Delta| e^{i\varphi/2}}{\gamma_B\xi_0\sqrt{\omega_n^2+|\Delta|^2}},\quad x=L
\\
(1\mp i\operatorname{sgn}(\omega_n)K\cos(2\alpha))\partial_y\tilde{f}_\pm\pm i\operatorname{sgn}(\omega_n)K\sin(2\alpha)\partial_x\tilde{f}_\pm&=0,\qquad \qquad \qquad \ \ \, y=0,\ L.
\end{align}
\end{subequations}

We solved the linearized Usadel equation, Eq.~(\ref{LinearUsadel}), together with these boundary conditions to obtain the results in Figs. \ref{SAM_Malpha0}-\ref{SAMS_zeropi} in the main text.
\end{document}